\documentclass[11pt, a4paper]{article}
\usepackage{amssymb}
\usepackage{amsmath}
\usepackage{graphicx}
\usepackage{subfigure}

\usepackage[authoryear]{natbib}
\usepackage{times,epsfig,makeidx,amsfonts,amsmath,dcolumn,enumerate,multirow,graphicx,amssymb,bm,color,amsthm}
\usepackage[margin=1in]{geometry}
\usepackage{algorithm}
\usepackage[]{algorithmic}
\usepackage{setspace,xspace}

\DeclareMathOperator{\supp}{supp}

\newcommand{\asconv}{\overset{a.s}{\rightarrow}}
\newcommand{\aseq}{\overset{a.s}{=}}
\newcommand{\asleq}{\overset{a.s}{\leq}}

\newtheorem{proposition}{Proposition}

\newtheorem{corollary}{Corollary}
\newtheorem{definition}{Definition}
\newtheorem{example}{Example}

\newcommand{\amj}[1]{\textcolor{blue}{#1}\xspace}
\renewcommand{\amj}[1]{#1\xspace}
\begin{document}
{\singlespace
\title{\amj{A Simple Approach to}\\Maximum Intractable Likelihood Estimation}

\author{F. J. Rubio and Adam M. Johansen\\\normalsize\texttt{\{F.J.Rubio,A.M.Johansen\}@warwick.ac.uk}\\{University of Warwick, Department of Statistics, Coventry, CV4 7AL, UK.}}
\maketitle}
\begin{abstract}
\singlespace
Approximate Bayesian Computation (ABC) can be viewed as an analytic
approximation of an intractable likelihood coupled with an elementary
simulation step. \amj{Such a view, combined with a suitable instrumental prior
  distribution permits} maximum-likelihood (or
maximum-a-posteriori) inference to be conducted, approximately, using
essentially the same techniques. \amj{An elementary} approach \amj{to this
  problem which simply obtains a nonparametric approximation of the likelihood
  surface which is then used as a smooth proxy for the likelihood in a
  subsequent maximisation step} is developed here and the convergence of this
class of algorithms is characterised theoretically. The use of non-sufficient
summary statistics {in this context} is considered. Applying
the proposed method to {four} problems demonstrates good
performance. The proposed approach provides an alternative for approximating
the maximum likelihood estimator (MLE) in complex scenarios.

\medskip

\end{abstract}

\section{Introduction}\label{intro}
Modern applied statistics must deal with many settings in which the pointwise
evaluation of the likelihood function, even up to a normalising constant, is
impossible or computationally infeasible.
Areas such as financial modelling,
genetics, geostatistics, {neurophysiology} and stochastic dynamical systems provide numerous examples
of this (see e.g. \citealp{CS54}; \citealp{Prit99}; and \citealp{T09}).
It is consequently difficult to perform any inference (classical or Bayesian)
about the parameters of the model. Various approaches
to overcome this difficulty have been proposed. For instance, Composite
Likelihood methods \citep{CR04}, for approximating the likelihood function,
and Approximate Bayesian Computational methods (ABC; \citealp{Prit99,B02}), for
approximating the posterior distribution, have been extensively studied in the
statistical literature. Here, we study the use of ABC methods, under an
appropriate choice of instrumental prior distribution, to approximate the
maximum likelihood estimator.

It is well-known that ABC produces a sample approximation of the
posterior distribution \citep{B02} in which there exists a \amj{deterministic}
approximation error in addition to Monte Carlo variability. {The quality of the approximation to the posterior and theoretical properties
of the estimators obtained with ABC have been studied in \cite{Wilkinson08};
\cite{Marin11}; \cite{Dean11} and \cite{Fear12}. The use of ABC posterior
samples for conducting model comparison was studied in \cite{Didelot11} and
\cite{Robert11b}.} Using this sample
approximation to characterise the mode of the posterior would in principle
allow (approximate) maximum {\sl a posteriori} (MAP) estimation.
Furthermore, using a uniform prior distribution, {under the parameterisation of interest}, over any set which contains
the MLE will lead to a MAP estimate which coincides with the MLE.
In low-dimensional problems if we have a sample from the posterior distribution
of the parameters, we can estimate its mode by using either nonparametric estimators
of the density {or another mode--seeking technique such as the
  \emph{mean-shift} algorithm \citep{FH75}}. Therefore, in contexts where the
likelihood function is intractable we can use these results to obtain an approximation of
the MLE. We will denote the estimator obtained with this approximation AMLE.

Although \cite{Marjoram03} noted that ``It [ABC] can also be used in
frequentist applications, in particular for maximum-likelihood estimation''
this idea does not seem to have been developed. A method based around
maximisation of a non-parametric estimate of the log likelihood function was
proposed by \cite{Diggle84} in the particular case of simple random samples;
their approach involved sampling numerous replicates of the data for each
parameter value and estimating the density in the data space. \cite{dV04}
proposes an importance sampling technique, rather closer in spirit
to the approach developed here, by which a smoothed kernel estimation
of the likelihood function up to a proportionality constant can be obtained
in the particular case of state space models provided that techniques for
sampling from the joint distribution of unknown parameters and latent states
are available --- not a requirement of the more general ABC technique developed below. The same
idea was applied and analysed in the context of the estimation of location
parameters, with particular emphasis on symmetric distributions, by \cite{Jaki08}.
The particular case of parameter estimation in hidden
Markov models was also investigated by \cite{Dean11}, who relied upon the specific
structure of that problem. To the best of our knowledge neither MAP estimation
nor maximum likelihood estimation in general, implemented directly via the
``ABC approximation'' combined with maximisation of an estimated density,
have been studied in the literature. {However, there has been a
  lot of interest in this type of problem using different approaches
  \citep{CK12,F12,M12} since we completed the first version of this work
  \citep{RJ12}.}

The estimation of the mode of nonparametric kernel density estimators which may seem, at first, to be a hopeless task has also received a
lot of attention (see  e.g. \citealp{P62,K73,R88,A03,BF06}). Alternative
nonparametric density estimators which could also be considered within the AMLE
context have been proposed recently in \cite{C10,J12}.

{The remainder of this paper is organised as follows}. In Section \ref{ABCapprox}, we present a brief description of ABC methods.
In Section \ref{sec:abcmle} we describe how to use these methods to approximate the MLE and present theoretical results to justify such use of ABC
methods. In Section \ref{examples}, we present
simulated and real examples to illustrate the use of the proposed MLE
approximation. Section \ref{sec:conclusion} concludes with a discussion of
both the developed techniques and the likelihood approximation obtained via
ABC in general.


\section{Approximate Bayesian Computation}\label{ABCapprox}

We assume throughout this and the following section that all distributions of
interest admit densities with  respect to an appropriate version of Lebesgue
measure, wherever this is possible, although this assumption can easily be
relaxed. {Let ${\bf x}=(x_1,...,x_n)\in{\mathbb R}^{q\times
    n}$} be a \amj{simple random} sample from a distribution $f(\cdot\vert \bm{\theta})$ with support contained in ${\mathbb R}^q$, $\bm{\theta}\in \bm{\Theta}\subset {\mathbb R}^d$; ${\mathcal L}(\bm{\theta};{\bf x})$ be the corresponding likelihood function, $\pi(\bm{\theta})$ be a prior distribution over the parameter $\bm{\theta}$ and $\pi(\bm{\theta}\vert{\bf x})$ the corresponding posterior distribution. Consider the following approximation to the posterior
\begin{eqnarray}\label{apppost}
\widehat{\pi}_{\varepsilon}(\bm{\theta}\vert{\bf x}) = \dfrac{\widehat{f}_{\varepsilon}({\bf x}\vert\bm{\theta})\pi(\bm{\theta})}{\int_{\Theta}\widehat{f}_{\varepsilon}({\bf x}\vert{\bf t})\pi({\bf t})d{\bf t}},
\end{eqnarray}
\noindent where
\begin{eqnarray}\label{applike}
\widehat{f}_{\varepsilon}({\bf x}\vert\bm{\theta}) = \int_{{\mathbb
    R}^n}K_{\varepsilon}({\bf x}\vert {\bf y}) f({\bf y}\vert\bm{\theta}) d{\bf{y}} ,
\end{eqnarray}
\noindent is an approximation of the likelihood function and
$K_{\varepsilon}({\bf x}\vert {\bf y})$ is a normalised Markov
kernel. \amj{$K_\varepsilon(\cdot|{\bf{y}})$ is typically concentrated around ${\bf
  y}$ with $\varepsilon$ acting as a scale parameter}. It's clear that (2) is
a smoothed version of the true likelihood and it has been argued that the
maximisation of such an approximation can in some circumstances lead to better performance
than the maximisation of the likelihood itself \citep{Ionides05}, providing an
additional motivation for the investigation of MLE via this approximation.
The approximation can be further motivated by noting that under weak
regularity conditions, the distribution
$\widehat{\pi}_{\varepsilon}(\bm{\theta}\vert{\bf x})$ is close (in some
sense) to the true posterior $\pi(\bm{\theta}\vert{\bf x})$ when $\varepsilon$
is sufficiently small. The simplest approach to ABC samples directly from
$(\ref{apppost})$ by the rejection sampling approach presented in Algorithm \ref{alg:abc}.

\begin{algorithm}[h!]
\begin{algorithmic}[1]
\STATE Simulate $\bm{\theta}^{\prime}$ from the prior distribution $\pi(\cdot)$.
\STATE Generate ${\bf y}$ from the model $f(\cdot\vert\bm{\theta}^{\prime})$.
\STATE Accept $\bm{\theta}^{\prime}$ with probability $\propto K_{\varepsilon}({\bf x} \vert {\bf y})$ otherwise return to step 1.
\end{algorithmic}
\caption{The basic ABC algorithm.}\label{alg:abc}
\end{algorithm}

Now, let $\bm{\eta}:{\mathbb R}^{n\cdot q} \rightarrow {\mathbb R}^m$ be a summary
statistic, $\rho:{\mathbb R}^m\times{\mathbb R}^m\rightarrow {\mathbb R}^+$ be a metric and
$\varepsilon>0$. The simplest ABC algorithm can be formulated in this way using the kernel

\begin{eqnarray}\label{kernelforABC}
K_{\varepsilon}({\bf x} \vert {\bf y}) \propto
\begin{cases} 1 & \text{if $\rho(\eta({\bf x}),\eta({\bf y}))<\varepsilon$,}
\\
0 &\text{otherwise.}
\end{cases}
\end{eqnarray}

The ABC rejection algorithm of \cite{Prit99} can be obtained simply by
setting $\eta({\bf x})={\bf x}$. Several improvements to the ABC method have
been proposed in order to increase the acceptance rate, see \cite{B02}, \cite{Marjoram03} and \cite{S07} for good surveys of these. An exhaustive summary of
these developments falls outside the scope of the present paper.

\section{Maximising Intractable Likelihoods}\label{sec:abcmle}
\subsection{Algorithm}\label{sec:amle}
 Point estimation of $\bm{\theta}$, by MLE and MAP estimation in particular,
 has been extensively studied \citep{LC98}. Recall that the MLE,
 $\widehat{\bm{\theta}}$, and the MAP estimator $\tilde{\bm{\theta}}$ are the
 values of $\bm\theta$ which maximise the likelihood or posterior density for the
 realised data.

These two quantities coincide when the prior distribution is constant (e.g. a
uniform prior $\pi(\bm{\theta})$ on a suitable (necessarily compact) set ${\bf
  D}$ which contains $\widehat{\bm{\theta}}$). Therefore, if we use a suitable
uniform prior, it is possible to approximate the MLE by using ABC methods to
generate an approximate sample from the posterior and then approximating the
MAP using this sample. In a different context in which the likelihood can be
evaluated pointwise, simulation-based MLEs which use a similar construction
have been shown to perform well (see, e.g., \citealp{GY03}, \citealp{L07} and \citealp{JDD08}). In the present setting the
optimisation step can be implemented by estimating the posterior density of
$\bm{\theta}$ using a nonparametric estimator (e.g. a kernel density
estimator) and then maximising this function: Algorithm \ref{alg:amle}.

\amj{We note that we have not here considered similar simulation-based approaches to the direct
optimisation of the likelihood function for a number of reasons. One is
computational cost (not having access to the likelihood even pointwise means
that distributions concentrated around the mode could be constructed only by
introducing several replicates of the data and the rejection or other
mechanism used to produce samples from this distribution will become
increasingly inefficient as the number of replicates increases);
another is that the proposed method has the additional advantages
that it fully characterises the likelihood surface and can be conducted
concurrently with Bayesian analysis with no additional simulation effort.}

\begin{algorithm}
\caption{The AMLE Algorithm}\label{alg:amle}
\begin{algorithmic}[1]
\STATE Obtain a sample $\bm{\theta}^*_{m,\varepsilon}=(\bm{\theta}^*_{m,\varepsilon,1},...,\bm{\theta}^*_{m,\varepsilon,m})$ from $\widehat{\pi}_{\varepsilon}(\bm{\theta}\vert{\bf x})$.
\STATE Using the sample $\bm{\theta}^*_{m,\varepsilon}$ construct a nonparametric estimator $\widehat\varphi$ of the density $\widehat{\pi}_{\varepsilon}(\bm{\theta}\vert{\bf x})$.
\STATE Calculate the maximum of $\widehat\varphi$, $\tilde{\bm{\theta}}_{m,\varepsilon}$. This is an approximation of the MLE $\widehat{\bm{\theta}}$.
\end{algorithmic}
\end{algorithm}

Note that the first step of this algorithm can be implemented rather
generally by using essentially any algorithm which can be used in the standard
ABC context. It is not necessary to obtain an iid sample from the distribution
$\widehat{\pi}_{\varepsilon}$: provided the sample is appropriate for
approximating that distribution it can in principle be employed in the AMLE
context (although correlation between samples obtained using MCMC techniques
and importance weights and dependence arising from the use of SMC can
complicate density estimation, it is not as problematic as might be
expected \citep{Skold03}).

A still more general algorithm could be implemented: using any prior
which has mass in some neighbourhood of the MLE and maximising the product of
the estimated likelihood and the reciprocal of this prior (assuming that the
likelihood estimate has lighter tails than the prior, not an onerous condition
when density estimation is used to obtain that estimate) will also provide an estimate of the
likelihood maximiser, an approach which was exploited by \cite{dV04} (who
provided also an analysis of the smoothing bias produced by this technique in
their context). In the interests of parsimony we do not pursue this
approach here, \amj{and throughout the remainder of this document we assume that a
uniform prior over some set $D$ which includes the MLE is used}, although we
note that such an extension eliminates the requirement that a compact set
containing a maximiser of the likelihood be identified in advance.

One obvious concern is that the approach could not be expected to work well
when the parameter space is of high dimension: it is well known that density
estimators in high-dimensional settings converge very slowly. Three things
mitigate this problem in the present context:
\begin{itemize}
\item  Many of the applications of ABC
have been to problems with extremely complex likelihoods which have only a
small number of parameters (such as {the examples} considered
below).
\item When the parameter space is of high dimension one could
  employ composite likelihood techniques with low-dimensional components
  estimated via AMLE. Provided appropriate parameter subsets are selected, the
  loss of efficiency will not be too severe in many
  cases. {Alternatively, a different \emph{mode-seeking}
    algorithm could be employed \citep{FH75}.}
\item In certain contexts, as discussed below Proposition \ref{prop:stats}, it
  may not be necessary to employ the density estimation step at all.
\end{itemize}

Finally, we note that direct maximisation of the smoothed likelihood
approximation (\ref{applike}) can be interpreted as a pseudo-likelihood technique \citep{Besag75}, with
the Monte Carlo component of the AMLE algorithm providing an approximation to this pseudo-likelihood.
\subsection{Asymptotic Behaviour}\label{results}
In this section we provide some theoretical results which justify the
approach presented in Section \ref{sec:amle} under similar conditions to
those used to motivate the standard ABC approach. \amj{We assume throughout that
the MLE exists in the model under consideration but that the likelihood is
intractable; in the case of non-compact parameter spaces, for example, this
may require verification on a case-by-case basis.}

We begin by showing pointwise convergence of the posterior (and hence
likelihood) approximation under reasonable regularity conditions. \amj{It is
convenient first to introduce the following concentration condition on the
class of ABC kernels which are employed:}

\begin{description}
\item[Condition K] A family of symmetric Markov kernels with
  densities $K_{\varepsilon}$  indexed by $\varepsilon > 0$ is said to satisfy
  the concentration condition provided that its members become
  increasingly concentrated as $\varepsilon$ decreases such that
\begin{eqnarray*}
\int_{{\mathcal B}_{\varepsilon}({\bf x})} K_{\varepsilon}({\bf x}\vert {\bf y}) d{\bf y} =
\int_{{\mathcal B}_{\varepsilon}({\bf x})} K_{\varepsilon}({\bf y}\vert {\bf x}) d{\bf y}
 =1,\,\,\, \forall \,\varepsilon>0.
\end{eqnarray*}
where ${\mathcal B}_\varepsilon({\bf x}) := \{ {\bf z} : |{\bf z} -{\bf x}|
\leq \varepsilon\}$.
\end{description}

\amj{As the user can freely specify $K$ this is not a problematic
  condition. It serves only to control the degree of smoothing which the ABC
  approximation of precision $\varepsilon$ can effect.}

\begin{proposition}\label{ABCSE}

{Let ${\bf x} =(x_1,...,x_n)\in{\mathbb R}^{q\times n}$} be a sample from a continuous distribution
$f(\cdot\vert\bm{\theta})$ with support contained in ${\mathbb R}^q$, $\bm{\theta}\in \Theta \subset{\mathbb R}^d$;
${\bf D} \subset{\mathbb R}^d$ be a compact set that contains
$\hat{\bm{\theta}}$, the MLE of $\bm{\theta}$; and let $K_{\varepsilon}$ be the
densities of a family of symmetric Markov kernels, \amj{which satisfies the
concentration condition (\textbf{K})}.

Suppose that
\begin{eqnarray*}
\sup_{({\bf t},\bm{\theta})\in{{\mathcal B}_{\epsilon}({\bf x})}\times {\bf D}} f({\bf t}\vert \bm{\theta})<\infty,
\end{eqnarray*}
\noindent for some $\epsilon > 0$. Then, for each $\bm{\theta}\in {\bf D}$
\begin{eqnarray*}
\lim_{\varepsilon \rightarrow 0}\widehat{\pi}_{\varepsilon}(\bm{\theta}\vert{\bf x})=\pi\left(\bm{\theta}\vert {\bf x}\right).
\end{eqnarray*}

\begin{proof}
It follows from the concentration condition that:
\begin{eqnarray*}
\widehat{f}_{\varepsilon}({\bf x}\vert\bm{\theta})=\int_{{\mathcal
    B}_{\varepsilon}({\bf x})} K_{\varepsilon}({\bf x}\vert{\bf y})f({\bf
  y}\vert \bm{\theta}) d{\bf y} .
\end{eqnarray*}
Furthermore, for each $\bm{\theta} \in {\bf D}$
\begin{eqnarray}\label{limit1}
\vert \widehat{f}_{\varepsilon}({\bf x}\vert\bm{\theta}) - f({\bf
  x}\vert\bm{\theta})  \vert \leq
\int_{{\mathcal
    B}_{\varepsilon}({\bf x})} d{\bf y} K_{\varepsilon}({\bf x}\vert{\bf y}) \left|f({\bf
  y}\vert \bm{\theta}) - f({\bf x}\vert \bm{\theta})\right|
\leq \sup_{{\bf y} \in{\mathcal
    B}_{\varepsilon}({\bf x})} \vert {f}({\bf y}\vert\bm{\theta}) - f({\bf x}\vert\bm{\theta})  \vert
\end{eqnarray}
which converges to 0 as $\varepsilon \rightarrow 0$ by continuity.
Therefore $\widehat{f}_{\varepsilon}({\bf x}\vert\bm{\theta}) \xrightarrow{\varepsilon\rightarrow 0} f({\bf x}\vert\bm{\theta})$.

Now, by bounded convergence (noting that boundedness of $\widehat{f}_\varepsilon({\bf x} \vert
\bm{\theta})$, for $\varepsilon < \epsilon$, follows from that of $f$ itself), we have that:
\begin{eqnarray}\label{limit2}
\lim_{\varepsilon\rightarrow 0}  \int_{\bf D} \widehat{f}_{\varepsilon}({\bf
  x}\vert\bm{\theta}^{\prime}) \pi({\bm{\theta}}^\prime)d\bm{\theta}^{\prime} =
\int_{\bf D} f({\bf x}\vert \bm{\theta}^{\prime}) \pi(\bm{\theta}^\prime)d\bm{\theta}^{\prime}.
\end{eqnarray}

The result follows by combining (\ref{limit1}) and (\ref{limit2}), whenever
$\pi({\bm \theta}|{\bf x})$ is itself well defined.
\end{proof}
\end{proposition}

This result can be strengthened by noting that it is straightforward to obtain
bounds on the error introduced at finite $\varepsilon$ \amj{if we assume Lipschitz
continuity of the true likelihood.}
Unfortunately, such conditions are not typically verifiable in problems of
interest. The following result, in which we show that whenever a sufficient
statistic is employed the \amj{ABC approximation converges pointwise to the
posterior distribution}, follows as a simple corollary to the previous
proposition. However, we provide an explicit proof based on a slightly
different argument in order to emphasize the role of sufficiency.

\begin{corollary}\label{ABCSE1}
Let ${\bf x} =(x_1,...,x_n)\in {\mathbb R}^{q\times n}$ be a sample from a distribution
$f(\cdot\vert\bm{\theta})$ over $\mathbb{R}^q$, $\eta:{\mathbb R}^{n\cdot
  q}\rightarrow{\mathbb R}^m$ be a sufficient statistic for
$\bm{\theta}\in\Theta\subset{\mathbb R}^d$, $\rho:{\mathbb R}^m\times{\mathbb
  R}^m \rightarrow {\mathbb R}_+$ be a metric and suppose that the density of
$\eta$, $f^{\eta}(\cdot\vert\bm{\theta})$, is $\rho-$continuous for 
every $\theta \in {\bf D}$. Let ${\bf D} \subset{\mathbb R}^d$ be a compact set, suppose that
\begin{eqnarray*}
\sup_{({\bf t},\bm{\theta})\in{{\mathcal B}_{\epsilon}}\times {\bf D}} f^{\eta}({\bf t}\vert \bm{\theta})<\infty,
\end{eqnarray*}
\noindent where ${\mathcal B}_{\epsilon}=\{{\bf t}\in{\mathbb R}^m: \rho(\eta({\bf x}),{\bf t})< \epsilon\}$ for some $\epsilon>0$ fixed. Then, for each $\bm{\theta}\in {\bf D}$ and the kernel $(\ref{kernelforABC})$
\begin{eqnarray*}
\lim_{\varepsilon \rightarrow 0}\hat{\pi}_{\varepsilon}(\bm{\theta}\vert{\bf x})=\pi\left(\bm{\theta}\vert {\bf x}\right).
\end{eqnarray*}

\proof
\amj{Using the integral Mean Value Theorem (as used in a similar context by
  \citet[Equation 6]{Dean11}) we find} that for $\bm{\theta}\in {\bf
  D}$ and any $\varepsilon \in (0,\epsilon)$:
\begin{eqnarray*}
\hat{f}_{\varepsilon}({\bf x}\vert\bm{\theta}) \propto \int I(\rho(\eta({\bf y}),\eta({\bf x}))<\varepsilon) f ({\bf y}\vert\bm{\theta}) d{\bf y}= \int_{{\mathcal B}_{\varepsilon}}f^{\eta}(\eta^{\prime}\vert\bm{\theta})d\eta^{\prime} = \lambda({\mathcal B}_{\varepsilon}) f^{\eta}\left(\xi(\bm{\theta},{\bf x}, \varepsilon)\vert\bm{\theta}\right),
\end{eqnarray*}
for some $\xi(\bm{\theta},{\bf x}, \varepsilon)\in {\mathcal
  B}_{\varepsilon}$, where $\lambda$ is the Lebesgue measure. Then
\begin{eqnarray*}
\hat{\pi}_{\varepsilon}(\bm{\theta}\vert{\bf
  x})=\frac{f^{\eta}\left(\xi(\bm{\theta},{\bf x}, \varepsilon) \vert
    {\bm{\theta}}\right) \pi({\bm \theta})}{\int_{\bf D}
  f^{\eta}\left(\xi(\bm{\theta}^{\prime},{\bf x},
    \varepsilon)\vert\bm{\theta}^{\prime}\right) \pi({\bm{\theta}^\prime})d\bm{\theta}^{\prime}}.
\end{eqnarray*}
As this holds for any sufficiently small $\varepsilon > 0$, we have by $\rho-$continuity of $f^{\eta}(\cdot\vert\bm{\theta})$:
\begin{eqnarray}\label{limit11}
\lim_{\varepsilon\rightarrow 0}  f^{\eta}\left(\xi( {\bm{\theta}}, {\bf x}, \varepsilon) \vert {\bm{\theta}}\right) = f^{\eta}\left({\eta}({\bf x}) \vert {\bm{\theta}}\right).
\end{eqnarray}
Using the Dominated Convergence Theorem we have
\begin{eqnarray}\label{limit21}
\lim_{\varepsilon\rightarrow 0}  \int_{\bf D}
f^{\eta}\left(\xi(\bm{\theta}^{\prime},{\bf x},
  \varepsilon)\vert\bm{\theta}^{\prime}\right)
\pi(\bm{\theta}^\prime)d\bm{\theta}^{\prime} = \int_{\bf D} f^{\eta}\left({\eta}({\bf
    x})\vert\bm{\theta}^{\prime}\right) \pi({\bm\theta}^\prime)d\bm{\theta}^{\prime}.
\end{eqnarray}
By the Fisher-Neyman factorization Theorem we have that there exists a
function $h:{\mathbb R}^{n\cdot q}\rightarrow{\mathbb R}_+$ such that
\begin{eqnarray}\label{NP1}
f\left({\bf x} \vert \bm{\theta}\right) = h({\bf x})f^{\eta}\left({\eta}({\bf x}) \vert {\bm{\theta}}\right)
\end{eqnarray}
The result follows by combining $(\ref{limit11})$, $(\ref{limit21})$ and $(\ref{NP1})$.$\square$
\end{corollary}


With only a slight strengthening of the conditions, Proposition \ref{ABCSE}
allows us to show convergence of the mode as $\varepsilon \rightarrow 0$ to
that of the true likelihood. It is known that pointwise convergence together
with equicontinuity on a  compact set implies uniform convergence
\citep{R76,W91}. Therefore, if in addition to the conditions in Proposition $\ref{ABCSE}$ we assume equicontinuity of $\widehat{\pi}_{\varepsilon}(\cdot \vert{\bf x})$ on ${\bf D}$, a rather weak additional condition, then the convergence to $\pi(\cdot\vert {\bf x})$ is uniform and we have the following direct corollary to Proposition \ref{ABCSE}:

\begin{corollary}\label{ConvModes}
Let $\widehat{\pi}_{\varepsilon}(\cdot\vert{\bf x})$ achieve its global
maximum at $\tilde{\bm{\theta}}_{\varepsilon}$ for each $\varepsilon>0$ and
suppose that $\pi(\cdot\vert {\bf x})$  has unique maximiser $\tilde{\bm{\theta}}$. Under the conditions in Proposition $\ref{ABCSE}$; if $\widehat{\pi}_{\varepsilon}(\cdot\vert{\bf x})$ is equicontinuous, then
\begin{eqnarray*}
\lim_{\varepsilon\rightarrow 0}\widehat{\pi}_{\varepsilon}(\tilde{\bm{\theta}}_{\varepsilon}\vert{\bf x}) = \pi(\tilde{\bm{\theta}}\vert {\bf x}).
\end{eqnarray*}
\end{corollary}


Using these results we can show that for \amj{a simple random} sample $\bm{\theta}^*_{m,\varepsilon}=(\bm{\theta}^*_{m,\varepsilon,1},...,\bm{\theta}^*_{m,\varepsilon,m})$ from the distribution $\widehat{\pi}_{\varepsilon}(\cdot\vert{\bf x})$ with mode at $\tilde{\bm{\theta}}_{\varepsilon}$ and an estimator $\tilde{\bm{\theta}}_{m,\varepsilon}$ , based on $\bm{\theta}^*_{m,\varepsilon}$, of $\tilde{\bm{\theta}}_{\varepsilon}$, such that $\tilde{\bm{\theta}}_{m,\varepsilon} \rightarrow
\tilde{\bm{\theta}}_{\varepsilon}$ almost surely when $m \rightarrow \infty$, we have that for any $\gamma>0$ there exists $\varepsilon>0$ such that
\begin{eqnarray*}
\lim_{m\rightarrow \infty}\left\vert \widehat{\pi}_{\varepsilon}\left(\tilde{\bm{\theta}}_{m,\varepsilon} \vert {\bf x} \right) -\pi\left(\tilde{\bm{\theta}} \vert {\bf x}\right)\right\vert \leq \gamma,\,\,\, \text{a.s.}
\end{eqnarray*}

That is, in the case of a sufficiently well-behaved density estimation
procedure, using the simple form of the ABC estimator (Algorithm
\ref{alg:abc}) we have that for any level of precision $\gamma$, the maximum of
the AMLE approximation will, for large enough ABC samples, almost surely be
$\gamma-$close to the maximum of the posterior distribution of interest, which
coincides with the MLE under the given conditions. A simple
continuity argument suffices to justify the use of
$\tilde{\bm{\theta}}_{m,\varepsilon}$ to approximate $\tilde{\bm{\theta}}$ for
large $m$ and small $\varepsilon$.

The convergence shown in the above results depends on the use of a sufficient
statistic. In contexts where the likelihood is intractable, this may not be
available. In the ABC literature, it has become common to employ summary
statistics which are not sufficient in this setting. Although it is possible
to characterise the likelihood approximation in this setting, it is difficult
to draw useful conclusions from such a characterisation. The construction of
appropriate summary statistics remains an active research area (see e.g.
\citealp{P10} and \citealp{Fear12}).

We finally provide one result which provides some support for the use of
certain non-sufficient statistics when there is a sufficient quantity of data
available. In particular we appeal to the large-sample limit in which it can
be seen that for a class of summary statistics the AMLE can almost
surely be made arbitrarily close to the true parameter value if a sufficiently small value of $\varepsilon$ can be used. This is, of course, an
idealisation, but provides some guidance on the properties required for
summary statistics to be suitable for this purpose and it provides some
reassurance that the use of such statistics can in principle lead to good
estimation performance. In this result we assume that the AMLE algorithm is
applied with the summary statistics filling the role of the data and hence the
ABC kernel is defined directly on the space of the summary statistics.

In order to establish this result, we require that, allowing $\eta_n({\bf x}) = \eta_n(x_1,\ldots,x_n)$ to denote a
sequence of $d_\eta$-dimensional summary statistics,
the following four conditions hold:

\begin{enumerate}[\bf {S}.i]
\item $\lim_{n\rightarrow\infty} \eta_n({\bf x})
\aseq g({\bm\theta})$ for
  $\pi-\text{a.e. } {\bm\theta}$
\item
$g:\Theta \rightarrow \mathbb{R}^{d_\eta}$ is an injective mapping. Letting $H = g(\mathbf{D}) \subset \mathbb{R}^{d_\eta}$  denote the image of the feasible
  parameter space under $g$, $g^{-1}:H
  \rightarrow \Theta$ is an $\alpha$-Lipschitz continuous function for some
  $\alpha \in \mathbb{R}_+$.
\item The ABC kernels, defined in the space of the summary statistics, satisfy
  condition {\bf K}, i.e. $K_\varepsilon^\eta(\cdot|\eta^\prime)$ it is concentrated
  within a ball of radius $\varepsilon$ for all $\varepsilon$: $\supp
  K_{\varepsilon}(\cdot|\eta^\prime) \subseteq
  \mathcal{B}_{\varepsilon}(\eta^\prime)$
\item The nonparametric estimator used always provides an estimate of
  the mode which lies within the convex hull of the sample.
\end{enumerate}

Some interpretation of these conditions seems appropriate. The first tells us
simply that the summary statistics converge to some function of the parameters
in the large sample limit, a mild requirement which is clearly necessary to
allow recovery of the parameters from the statistics. The second condition
strengthens this slightly, requiring that the limiting values of the
statistics and parameters exist in one-to-one correspondence and that this
correspondence is regular in a Lipschitz-sense. The remaining conditions
simply characterise the behaviour of the ABC approximation and the AMLE algorithm.

\begin{proposition}\label{prop:stats}
Let ${\bf x}=(x_1,x_2,\ldots)$ denote a sample with joint measure
$\mu(\cdot|{\bm\theta})$ for some ${\bm\theta} \in \mathbf{D}
\subset \Theta$. Let $\pi({\bm\theta})$ denote a prior density over
$\mathbf{D}$. Let $\eta_n({\bf x}) = \eta_n(x_1,\ldots,x_n)$ denote a
sequence of $d_\eta$-dimensional summary statistics with distributions $\mu^{\eta_n}(\cdot|{\bm\theta})$. Allow $\eta_n^\star$ to denote an observed
value of the sequence of statistics obtained from the model with ${\bm\theta}={\bm\theta}^\star$.

\medskip\noindent Assume that conditions \textbf{S.i}--\textbf{S.iv} hold.
Then:
\begin{enumerate}[(a)]
\item $\supp\lim_{n\rightarrow\infty} \pi_\varepsilon({\bm\theta}|\eta_n^\star)
  \subseteq \mathcal{B}_{\alpha\varepsilon}({\bm\theta}^\star)$ for
  $\mu(\cdot|{\bm\theta}^\star)$-almost every $\eta^\star$ for $\pi$-almost every ${\bm\theta}^\star$.
\item The AMLE approximation of the MLE lies within
  $\mathcal{B}_{\alpha\varepsilon}({\bm\theta}^\star)$ almost surely.
\end{enumerate}

\begin{proof}
Allowing $f_\varepsilon^{\eta_n}(\eta|{\bm\theta})$ to denote the ABC approximation
of the density of $\eta_n$ given ${\bm\theta}$, we have:
\begin{align*}
\lim_{n\rightarrow\infty} f_\varepsilon^{\eta_{n}}(\eta|{\bm\theta})
=
\lim_{n\rightarrow\infty} \int \mu^{\eta_n}(d\eta^\prime|{\bm\theta}) K_\varepsilon(\eta|\eta^\prime)
\aseq
\int \delta_{g({\bm\theta})}(d\eta^\prime) K_{\varepsilon}(\eta|\eta^\prime) = K_\varepsilon(\eta|g({\bm\theta}))
\end{align*}
where with the final equality following from \textbf{S.i} (noting that
almost sure convergence of $\eta_n$ to $g({\bm\theta})$ implies convergence in
distribution of $\eta_n$ to a degenerate random variable taking the value $g({\bm\theta}))$.

From which it is clear that $\supp \lim_{n \rightarrow \infty} f_\varepsilon^{\eta_n}(\cdot|{\bm\theta}) \subseteq
\mathcal{B}_{\varepsilon}(g({\bm\theta}))$ by \textbf{S.iii}.

And the ABC approximation to the posterior density
of ${\bm\theta}$, $\lim_{n\rightarrow\infty}\pi_\varepsilon(\cdot|\eta_n)$, may be
similarly constrained:
\begin{align*}
\lim_{n\rightarrow\infty} \pi_\varepsilon({\bm\theta}|\eta_n) > 0 &\Rightarrow &
\lim_{n\rightarrow\infty} || \eta_n - g({\bm\theta})|| \asleq& \varepsilon &
&\Rightarrow& \lim_{n\rightarrow\infty} || g^{-1}(\eta_n) - {\bm\theta}|| \asleq& \alpha\varepsilon
\end{align*}
using \textbf{S.ii}. And by assumption \textbf{S.i}, \textbf{S.ii} and the continuous mapping theorem we have that $g^{-1}(\eta_n^\star) \asconv
{\bm\theta}^\star$ giving result (a); result (b) follows immediately from \textbf{S.iv}.
\end{proof}
\end{proposition}

It is noteworthy that this proposition suggests that, at least in the
large sample limit, one can use any estimate of the mode which lies within the
convex hull of the sampled parameter values. The posterior mean would satisfy
this requirement and thus for large enough data sets it is not necessary to
employ the nonparametric density estimator at all in order to implement
AMLE. This is perhaps an unsurprising result and seems a natural consequence
of the usual Bayesian consistency results but it does have implications for
implementation of AMLE in settings with large amounts of data for which the
summary statistics are with high probability close to their limiting values.

\begin{example}[Location-Scale Families and Empirical Quantiles]
Consider a simple random sample from a location-scale family, in which we can write the distribution
functions in the form:
\begin{align*}
F(x_i|\mu,\sigma) = F_0((x_i - \mu)/\sigma)
\end{align*}
Allow $\eta_n^1 = \hat{F}^{-1}(q_1)$ and $\eta_n^2 = \hat{F}^{-1}(q_2)$ to
denote to empirical quantiles. By the Glivenko-Cantelli theorem, these
empirical quantiles converge almost-surely to the true quantiles:
\begin{align*}
\lim_{n\rightarrow\infty} \left(
\begin{array}{c}
\eta_n^1 \\
\eta_n^2
\end{array}
\right)
\aseq
\left(
\begin{array}{c}
F^{-1}(q_1|\mu,\sigma) \\
F^{-1}(q_2|\mu,\sigma)
\end{array}
\right)
\end{align*}
In the case of the location-scale family, we have that:
\begin{align*}
F^{-1}(q^i|\mu,\sigma) = \sigma F_0^{-1}(q^i) + \mu
\end{align*}
and we can find explicitly the mapping $g^{-1}$:
\begin{align*}
g^{-1}(\eta_n^1,\eta_n^2) = \left(
\begin{array}{c}
\frac{\eta_n^1 - \eta_n^2}{F_0^{-1}(q_1) - F_0^{-1}(q_2)} \\
\eta_n^1 -
\frac{\eta_n^1 - \eta_n^2}{F_0^{-1}(q_1) - F_0^{-1}(q_2)} F_0^{-1}(q_1)
\end{array}
\right)
\asconv
\left(
\begin{array}{c}
\sigma\\
\mu
\end{array}
\right)
\end{align*}
provided that $F_0^{-1}(q_1) \neq F_0^{-1}(q_2)$ which can be assured if $F_0$
is strictly increasing and $q_1 \neq q_2$. In this case we even obtain an
explicit form for $\alpha$.
\end{example}

\subsection{Use of kernel density estimators}
In this section we demonstrate that the simple Parzen estimator can be
employed within the AMLE context with the support of the results of the
previous section.

\begin{definition}\citep{P62}
Consider the problem of estimating a density with support on ${\mathbb R}^n$ from $m$ independent random vectors $({\bf Z_1},...,{\bf Z_m})$. Let $K$ be a kernel, $h_m$ be a bandwidth such that $h_m\rightarrow 0$ when $m\rightarrow \infty$, then a kernel density estimator is defined by

\begin{eqnarray*}
\widehat{\varphi}_m({\bf z}) = \dfrac{1}{m h_m^n}\sum_{j=1}^m K\left(\dfrac{{\bf z}-{\bf Z_j}}{h_m}\right).
\end{eqnarray*}
\end{definition}

Under the conditions $h_m \rightarrow 0$ and $m h_m^n/\log(m)\rightarrow
\infty$ together with Theorem $1$ from \cite{A03}, we have that
$\tilde{\bm{\theta}}_m \xrightarrow{a.s.} \tilde{\bm{\theta}}$ as
$m\rightarrow \infty$. Therefore, the results presented in the previous
section apply to the use of kernel density estimation. This demonstrates that this simple non-parametric estimator is adequate for approximation of the MLE via the AMLE strategy, at least asymptotically.

This is, of course, just one of many ways in which the density could be
estimated and more sophisticated techniques could very easily be employed and
justified in the AMLE context.

\section{Examples}\label{examples}

{We present four examples in increasing order of complexity. The first two examples illustrate the performance of the algorithm
in simple scenarios in which the solution is known;} \amj{the third compares
  the algorithm with a quantile-based method in a setting which has recently
  been studied using ABC and the final example demonstrates
performance on a challenging estimation problem which has recently attracted
some attention in the literature.}
In all the examples the simple ABC rejection algorithm was used, together with ABC kernel $(\ref{kernelforABC})$. For the second, third and fourth
examples, kernel density estimation is conducted using the R command `kde'
together with the bandwidth matrix obtained via the smoothed cross validation
approach of \cite{D05} using the command `Hscv' from the R package `ks'
\citep{D11}. R source code for these examples is available from the first
author upon request.

\subsection{Binomial Model}


Consider a sample of size $30$ simulated from a Binomial$(10,0.5)$ with
$\bar{x}=5.53$. Using the prior $\theta\sim \text{Unif}(0,1)$, a tolerance
$\varepsilon=0.1$, a sufficient statistic $\eta({\bf x})=\bar x$ and the Euclidean metric we simulate an ABC sample of size $10,000$ which, together with Gaussian kernel estimation of the posterior, gives the AMLE ${\tilde \theta}=0.552$.

\begin{figure}
\begin{center}
\subfigure[$D = (0.45,0.65)$]{
\psfig{figure=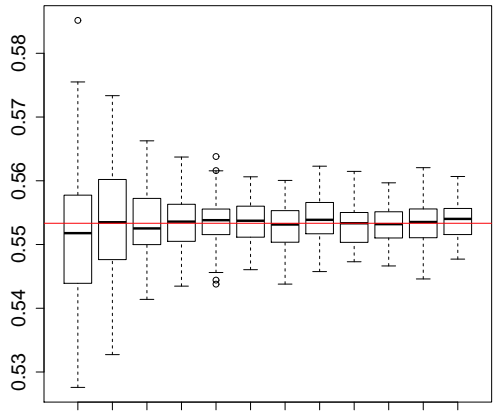, width=0.45 \textwidth}
}
\subfigure[$D=(0,1)$]{
\psfig{figure=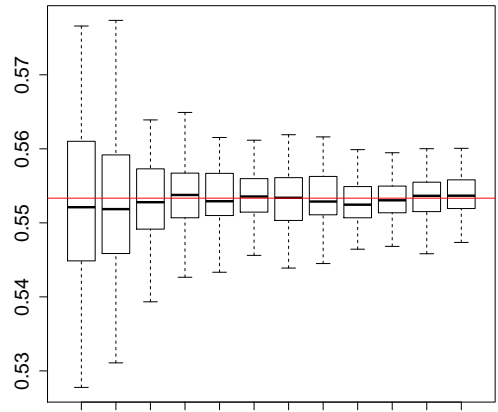, width=0.45 \textwidth}
}
\end{center}
\caption[]{Effect of $n \in \{30,100,1000,2000,...,10000\}$ for $\varepsilon=0.05$. The continuous red line represents the true MLE value.}
\label{fig:effectn}
\end{figure}

\begin{figure}
\begin{center}
\begin{tabular}{cc}
\psfig{figure=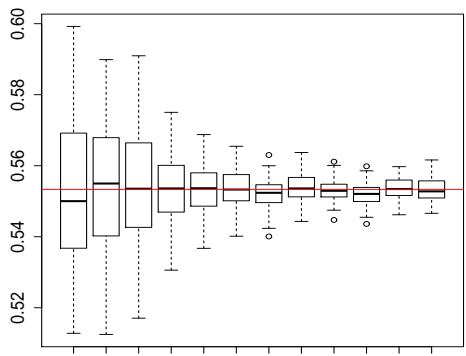, height=6.0cm, width=6.75cm} & %
\psfig{figure=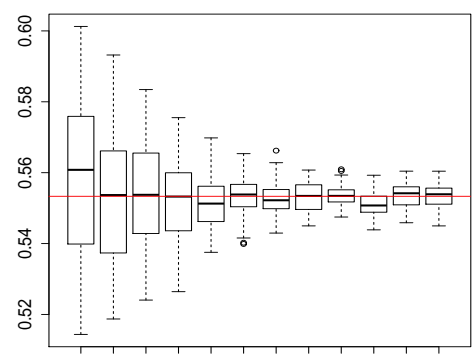, height=6.0cm, width=6.75cm} \\
(a) & (b)%
\end{tabular}
\end{center}
\caption[]{ \small Effect of $\varepsilon \in \{1,0.9,...,0.1,0.05,0.01\}$ for $n=10000$: (a) $D=(0.45,0.65)$; (b) $D=(0,1)$. The continuous red line represents the true MLE value}
\label{fig:effecteps}
\end{figure}

There are three quantities affecting the precision in the estimation of
${\widehat \theta}$: $D$, $n$ and $\varepsilon$. Figure $\ref{fig:effectn}$
illustrates the effect of varying $n \in
\{30,100,1000,2000,...,10000\}$ for a fixed $\varepsilon$, two different
choices of $D$ and an ABC sample of size $10,000$. Boxplots were obtained
using $100$ replications of the (stochastic) AMLE algorithm.
\amj{This demonstrates that although, unsurprising the acceptance rate and hence
computational efficienct is improved when some $D$ which is relatively
concentrated around the MLE is available, estimation precision remains good
when the full support of the parameter space is included in $D$ albeit at
greater computational cost}
(the choice $D=(0.45,0.65)$ produces an acceptance rate about 5 times greater than the choice $D=(0,1)$). Figure $\ref{fig:effecteps}$ shows the effect of $\varepsilon \in \{1,0.9,...,0.1,0.05,0.01\}$ for a fixed $n$ and two different choices of $D$. In this case we can note that the effect of $\varepsilon$ on the precision is significant. Again, the choice of $D$ affects only the acceptance rate.

\pagebreak

\subsection{Normal Model}

Consider a sample of size $100$ simulated from a Normal$(0,1)$ with sample
mean $\bar {\bf x}= -0.005$ and sample variance $s^2=1.004$. Suppose that both
parameters $(\mu,\sigma)$ are unknown. The MLE of $(\mu,\sigma)$ is simply
$(\widehat\mu,\widehat\sigma)=(-0.005,1.002)$.

\begin{figure}
\begin{center}
\begin{tabular}{cc}
\psfig{figure=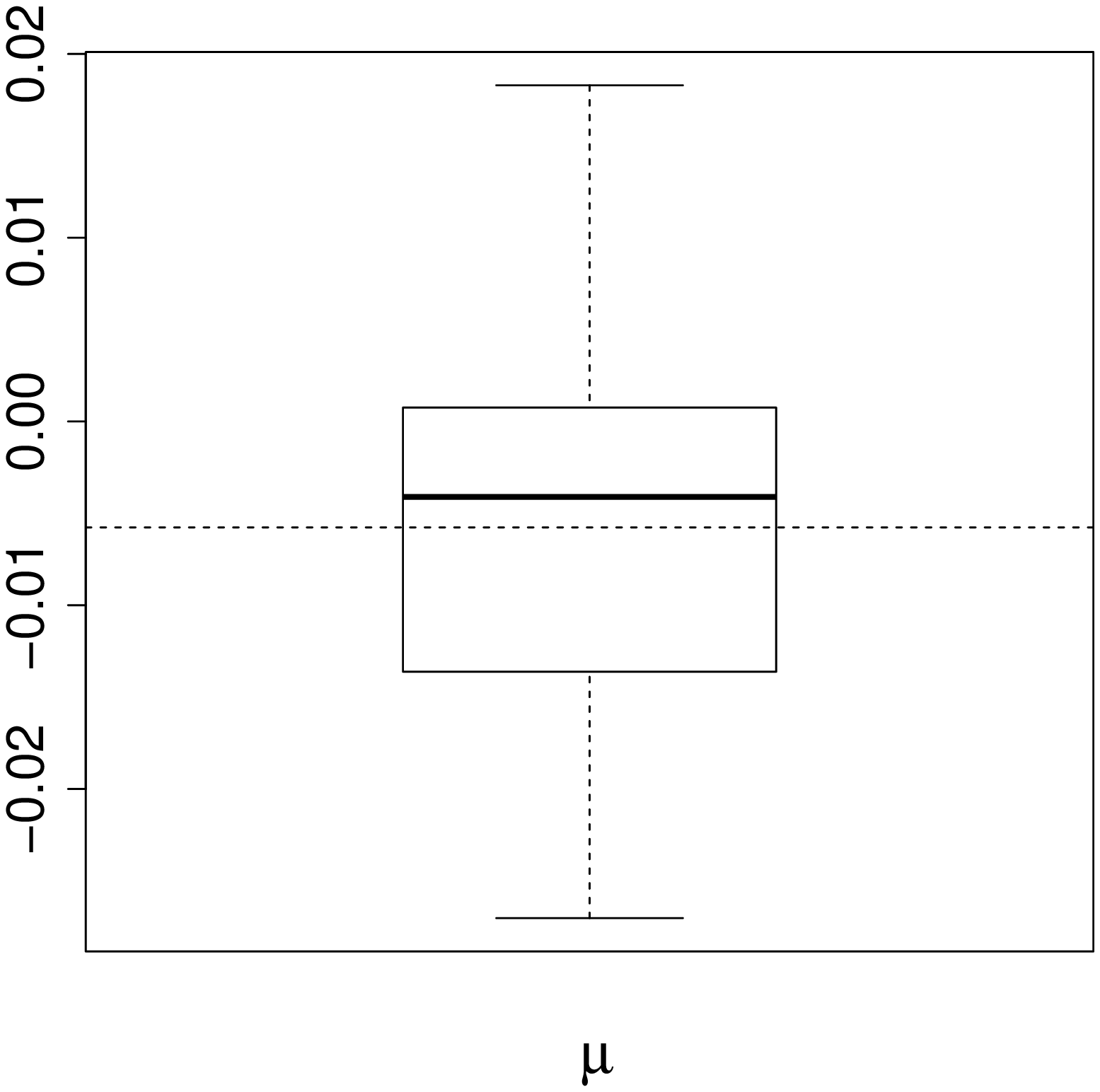, height=4.5cm, width=4.5cm} & %
\psfig{figure=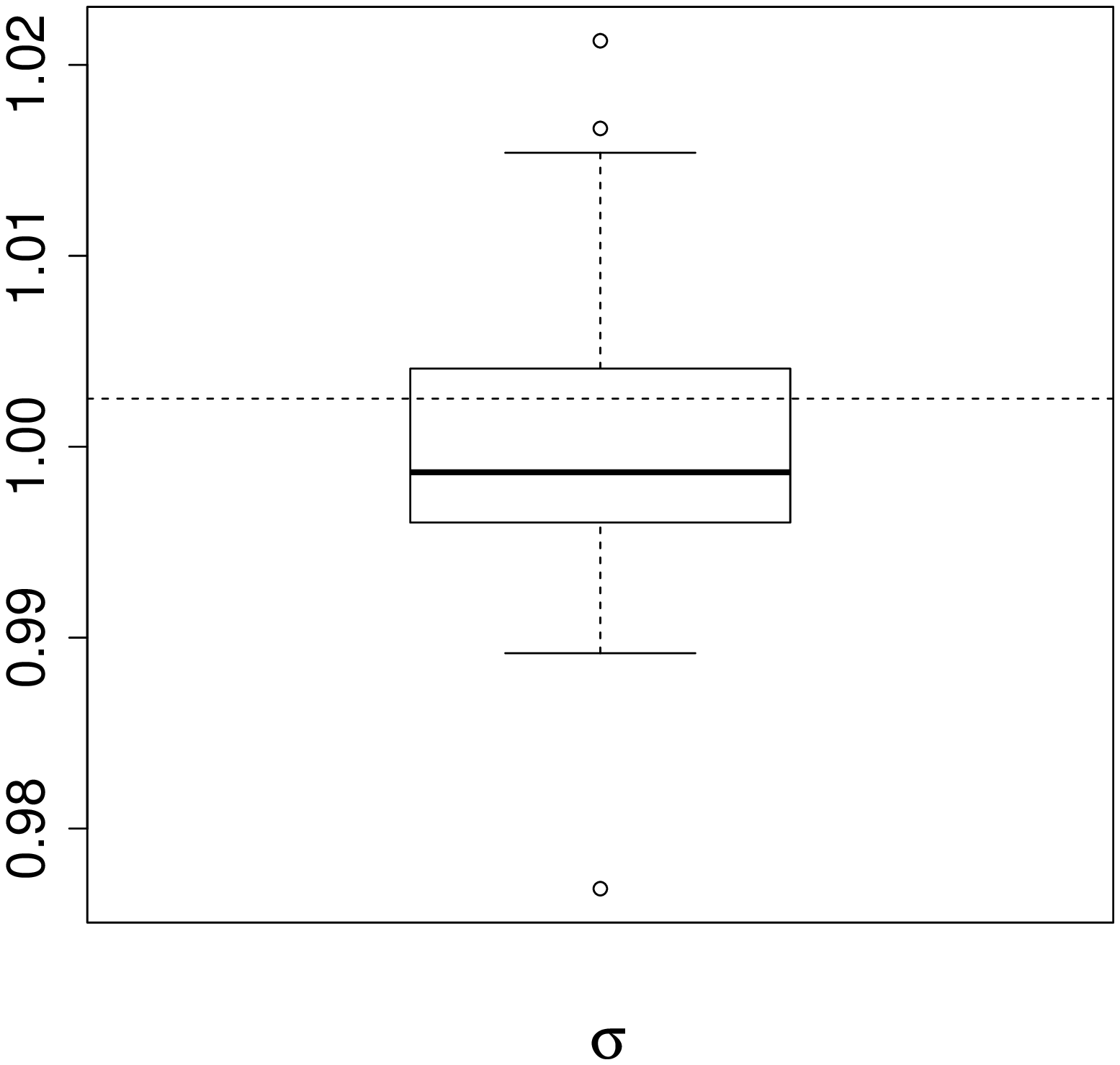, height=4.5cm, width=4.5cm} \\
(a) & (b) %
\end{tabular}
\end{center}
\caption{ \small Monte Carlo variability of the AMLE: (a) $\mu$; (b) $\sigma$. The dashed lines represent the true MLE value.}
\label{fig:MCN}
\end{figure}

Consider the priors $\mu\sim \text{Unif}(-0.25,0.25)$ and $\sigma\sim
\text{Unif}(0.75,1.25)$ \amj{(crude estimates of location and scale can often be
obtained from data, justifying such a choice; using broader prior support here
increases computational cost but does not prevent good estimation)}, a
tolerance $\varepsilon=0.01$, a sufficient statistic
$\eta({\bf x})=(\bar {\bf x}, s)$, the Euclidean metric, an ABC sample
of size $5,000$, and Gaussian kernel estimation of the posterior. Figure
$\ref{fig:MCN}$ illustrates Monte Carlo variability of the AMLE of
$(\mu,\sigma)$. Boxplots were obtained using $50$
replicates of the algorithm.

\subsection{{Financial application}}

Logarithmic daily return prices are typically modelled using L{\'e}vy
processes. For this reason, it is necessary to model the increments
(logarithmic returns) using an infinitely divisible distribution. It has been
found empirically that these observations have tails heavier than those of the
normal distribution, and therefore an attractive option is the use of the
$4-$parameter $(\alpha,\beta,\mu,\sigma)$ $\alpha-$stable family of
distributions, which can account for this behaviour. It is well known that
maximum likelihood estimation for this family of distributions is difficult. Various numerical approximations of the MLE have been proposed (see e.g. \citealp{M86,N01}). From a Bayesian perspective, \cite{P10} proposed the use of ABC methods to obtain an approximate posterior sample of the parameters. They propose six summary statistics that can be used for this purpose.

\begin{figure}[h!]
\begin{center}
\begin{tabular}{cc}
\psfig{figure=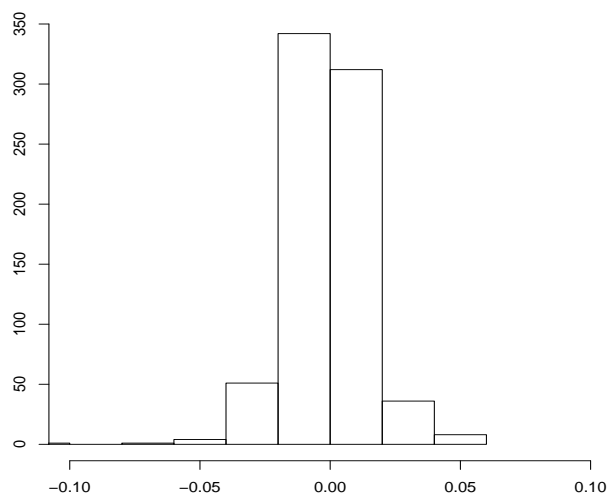, height=8.0cm, width=9cm}
\end{tabular}
\end{center}
\caption[]{ \small Logarithmic daily returns using the closing price of IBM ordinary stock Jan. 1 2009 to Jan. 1 2012.}
\label{fig:LR}
\end{figure}

Here, we analyse the logarithmic daily returns using the closing price of IBM
ordinary stock from Jan. 1 2009 to Jan. 1 2012. Figure $\ref{fig:LR}$ shows
the corresponding histogram. For this data set, the MLE using McCulloch's
quantile method implemented in the R package `fBasics' \citep{W10} is
$(\hat\alpha,\hat\beta,\hat\mu,\hat\sigma)$$ =(1.4930, -0.0780,-0.0007,  0.0073)$. 

Given the symmetry observed and in the spirit of parsimony, we consider the
skewness parameter $\beta$ to be $0$ in order to calculate the AMLE of the
parameters $(\alpha,\mu,\sigma)$. Based on the interpretation of these
parameters (shape, location and scale) and the data we use the priors
\begin{eqnarray*}
\alpha \sim U(1,2),\,\,\,\mu \sim U(-0.1,0.1),\,\,\, \sigma \sim U(0.0035,0.0125)
\end{eqnarray*}
\amj{which due to the scale of the data may appear concentrated but are, in fact, rather uninformative, allowing a location parameter essentially
anywhere within the convex hull of the data, scale motivated by similar
considerations and any value of the shape parameter consistent with the problem at hand}.

\begin{figure}[h!]
\begin{center}
\begin{tabular}{ccc}
\psfig{figure=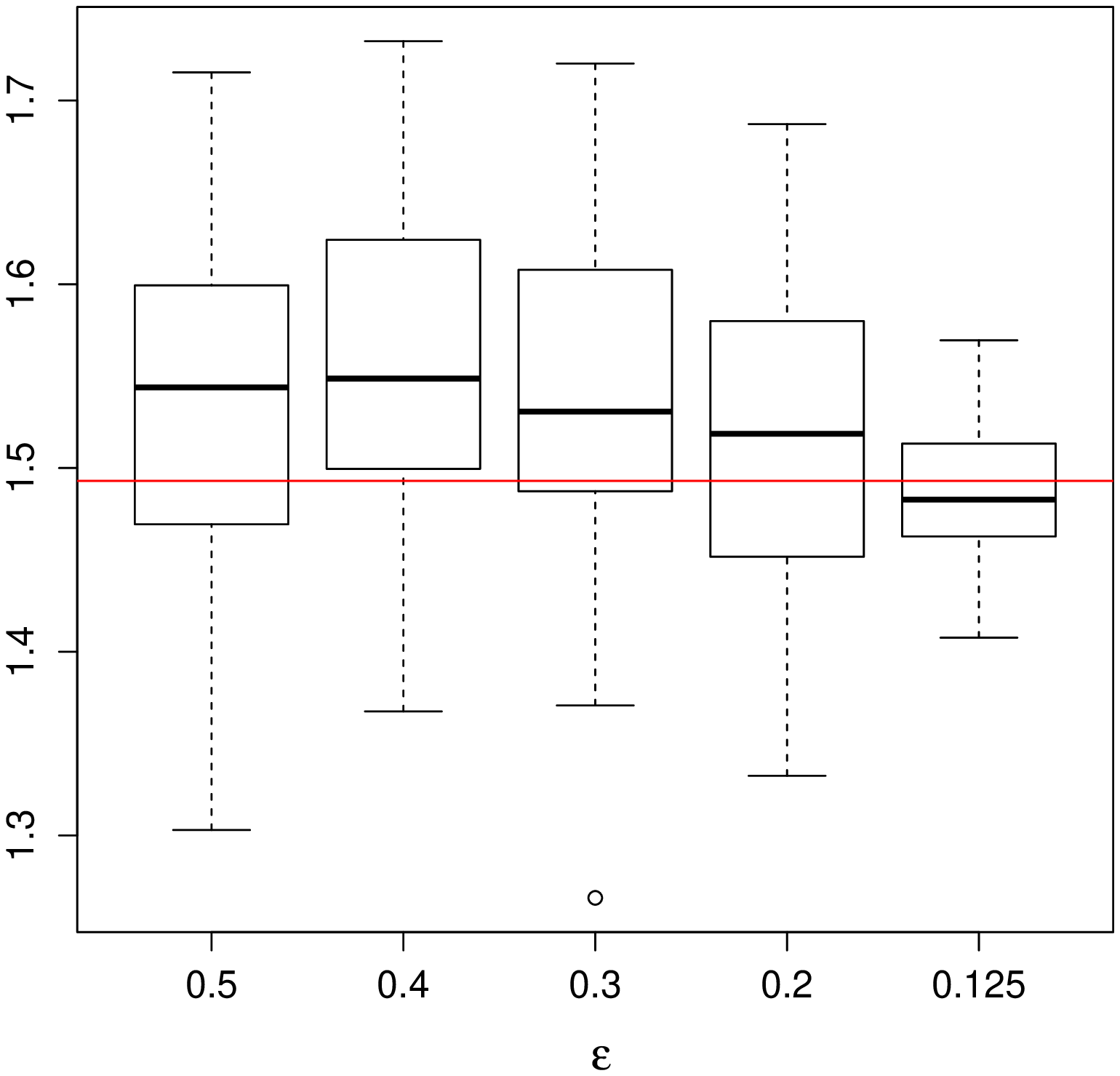, height=4.5cm, width=4.5cm} & %
\psfig{figure=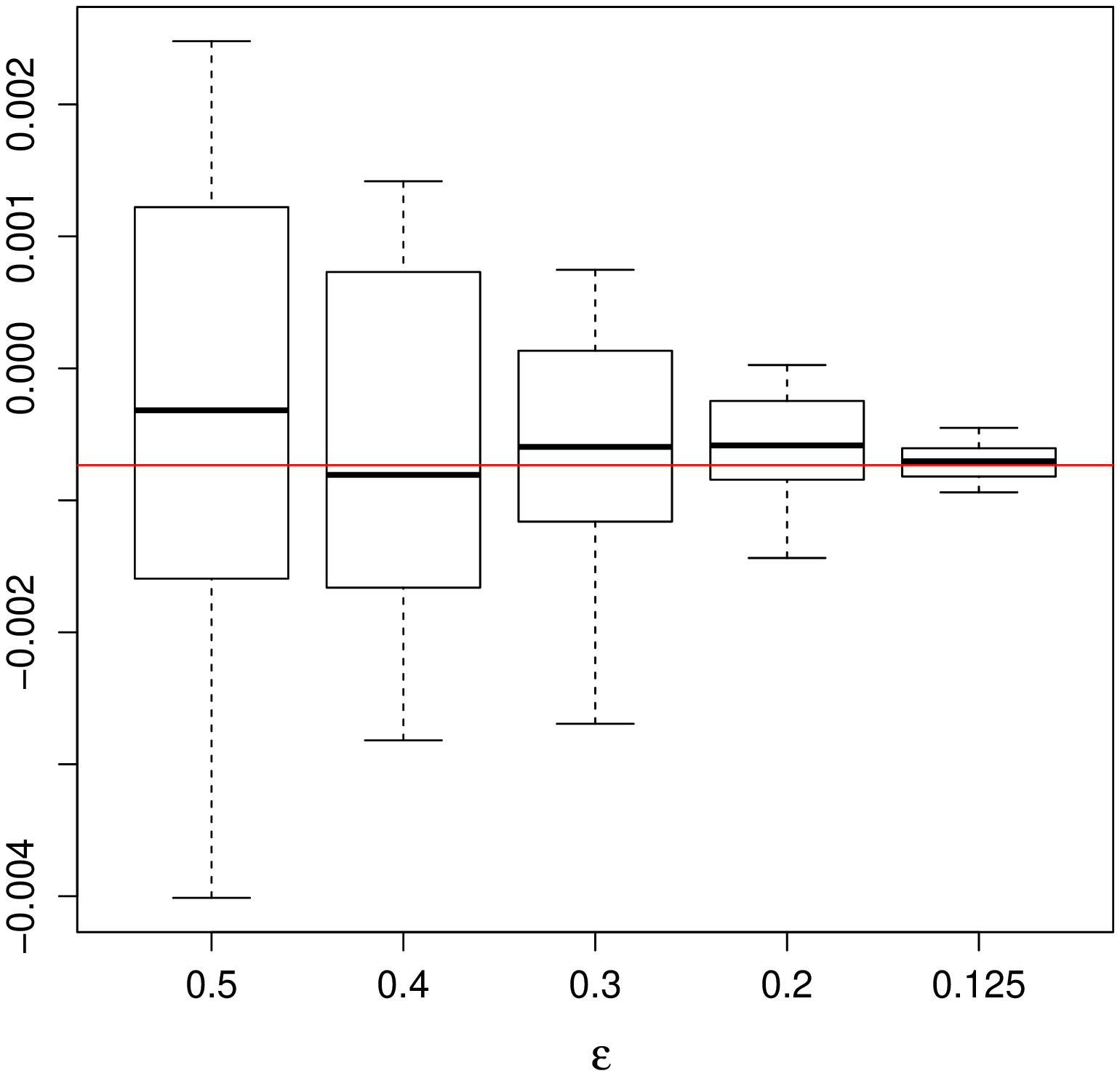, height=4.5cm, width=4.5cm} &
\psfig{figure=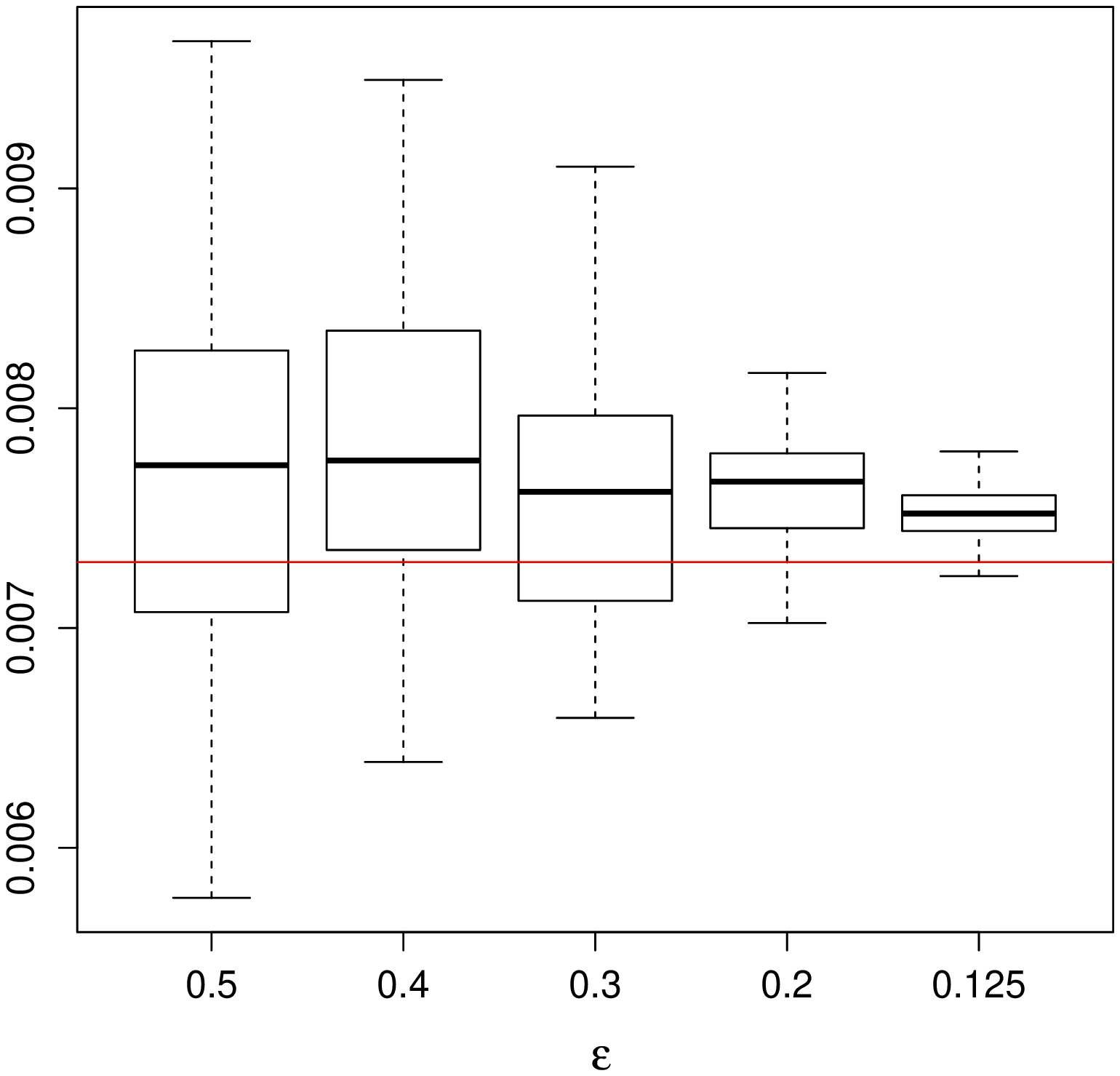, height=4.5cm, width=4.5cm} \\
(a) & (b) & (c)%
\end{tabular}
\end{center}
\caption{ \small Monte Carlo variability of the AMLE: (a) $\alpha$; (b) $\mu$;
  (c) $\sigma$. Horizontal lines represent McCulloch's estimator produced by the R package `fBasics'.}
\label{fig:MCLR}
\end{figure}

For the (non-sufficient) summary statistic we use proposal $S_4$ of \cite{P10}, which
consists of the values of the empirical characteristic function evaluated on an
appropriate grid. We use the grid $t\in\{$-250, -200, -100, -50, -10, 10, 50,
100, 200, 250$\}$, an ABC sample of size $2,500$, a tolerance $\varepsilon\in\{0.5,0.4,0.3,0.2,0.125\}$ and Gaussian kernel density
estimation. Figure $\ref{fig:MCLR}$ illustrates Monte Carlo variability of the
AMLE of $(\alpha,\mu,\sigma)$. Boxplots were obtained using $50$ replicates of
the AMLE procedure. A simulation study considering several simulated data sets
produced with common parameter values (results not shown) suggest that the
sampling variability in McCulloch's estimator exceeds the difference between
that estimator and the AMLE based upon $S_4$. In general,
considerable care must of course be taken in the selection of statistics ---
it is noteworthy that the quantiles used in McCulloch's own estimator satisfy
most of the requirements of Proposition \ref{prop:stats}, although it is not clear that it is
possible to demonstrate the Lipschitz-continuity of $g^{-1}$ in this case.

\pagebreak

\subsection{{Superposed gamma point processes}}

The modelling of an unknown number of superposed gamma point processes
provides another scenario with intractable likelihoods \amj{which is currently
  attracting some attention} \citep{CK12,M12}. Intractability of the
likelihood in this case is \amj{a consequence of the dependency between} the observations, which complicates the construction of their joint distribution. Superposed point processes have applications in a variety of areas, for instance \cite{CS54} present an application of this kind of processes in the context of neurophysiology. In this example we consider a simulated sample of size $88$ of $N=2$ superposed point processes with inter-arrival times identically distributed as a gamma random variable with shape parameter $\alpha=9$ and rate parameter $\beta=1$ observed in the interval $(0,t_0)$, with $t_0=420$. This choice is inspired by the simulated example presented in \cite{CK12}.

\begin{figure}[h!]
\begin{center}
\begin{tabular}{cc}
\psfig{figure=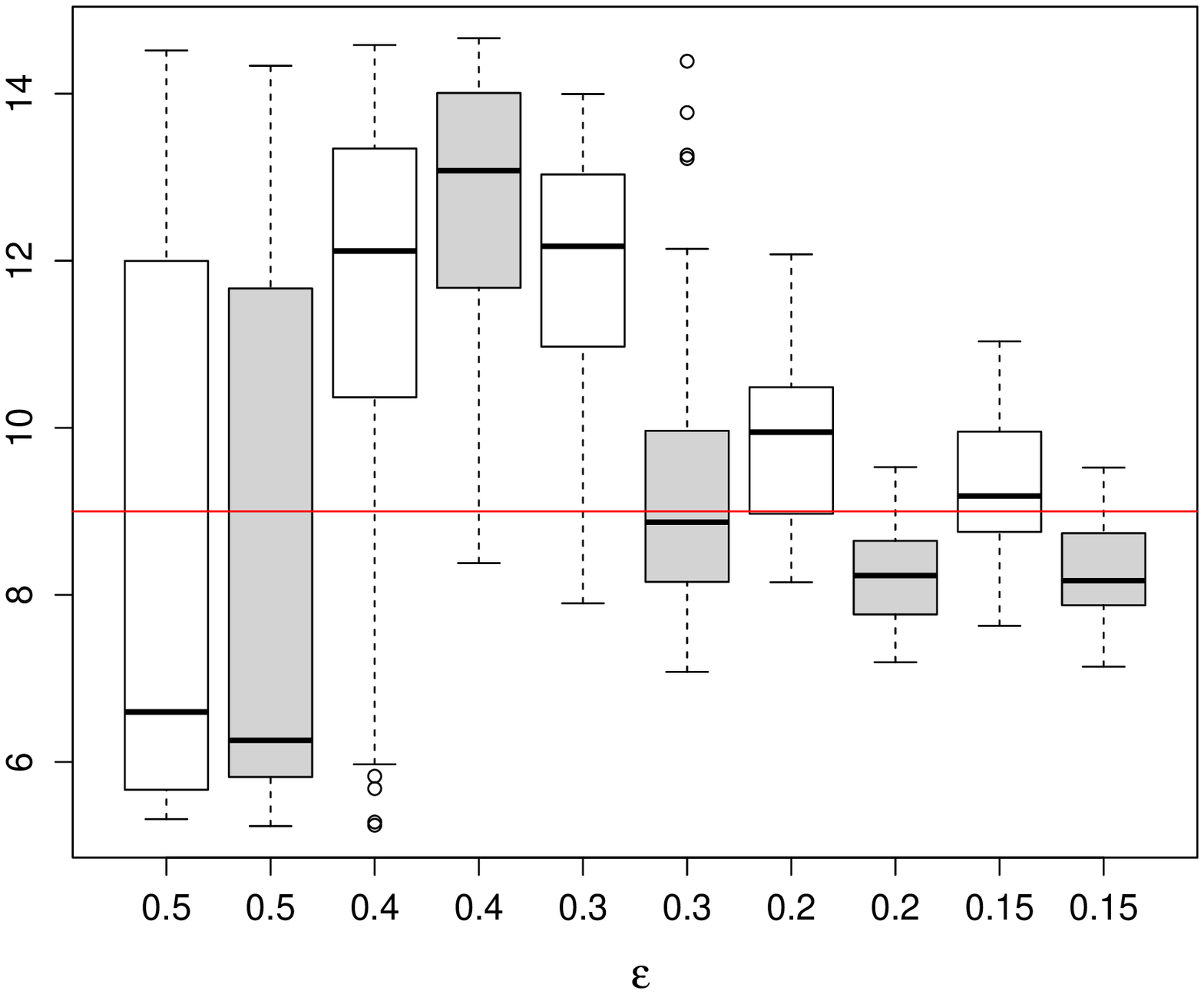, height=5cm, width=6cm} & %
\psfig{figure=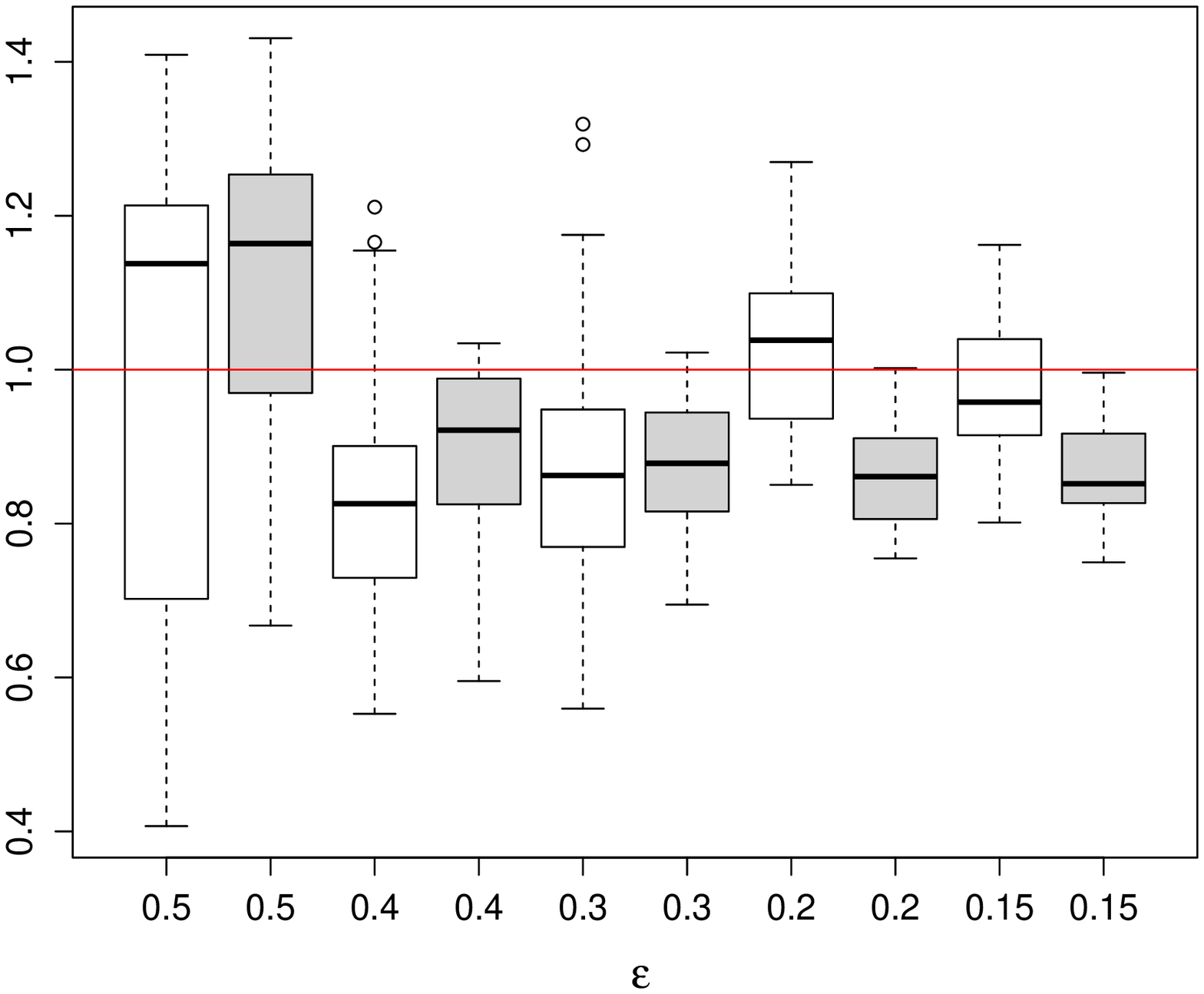, height=5cm, width=6cm} \\
(a) & (b) %
\end{tabular}
\end{center}
\caption[]{ \small Effect of $\varepsilon \in \{0.5,0.4,0.3,0.2,0.15\}$ for $n=5000$: (a) $\alpha$; (b) $\beta$. The AMLE samples with $8$ and $9$ summary statistics are presented in white and gray boxplots, respectively. The continuous red line represents the true value of the parameter.}
\label{fig:effectepsRP}
\end{figure}

\begin{figure}[h!]
\begin{center}
\begin{tabular}{cc}
\psfig{figure=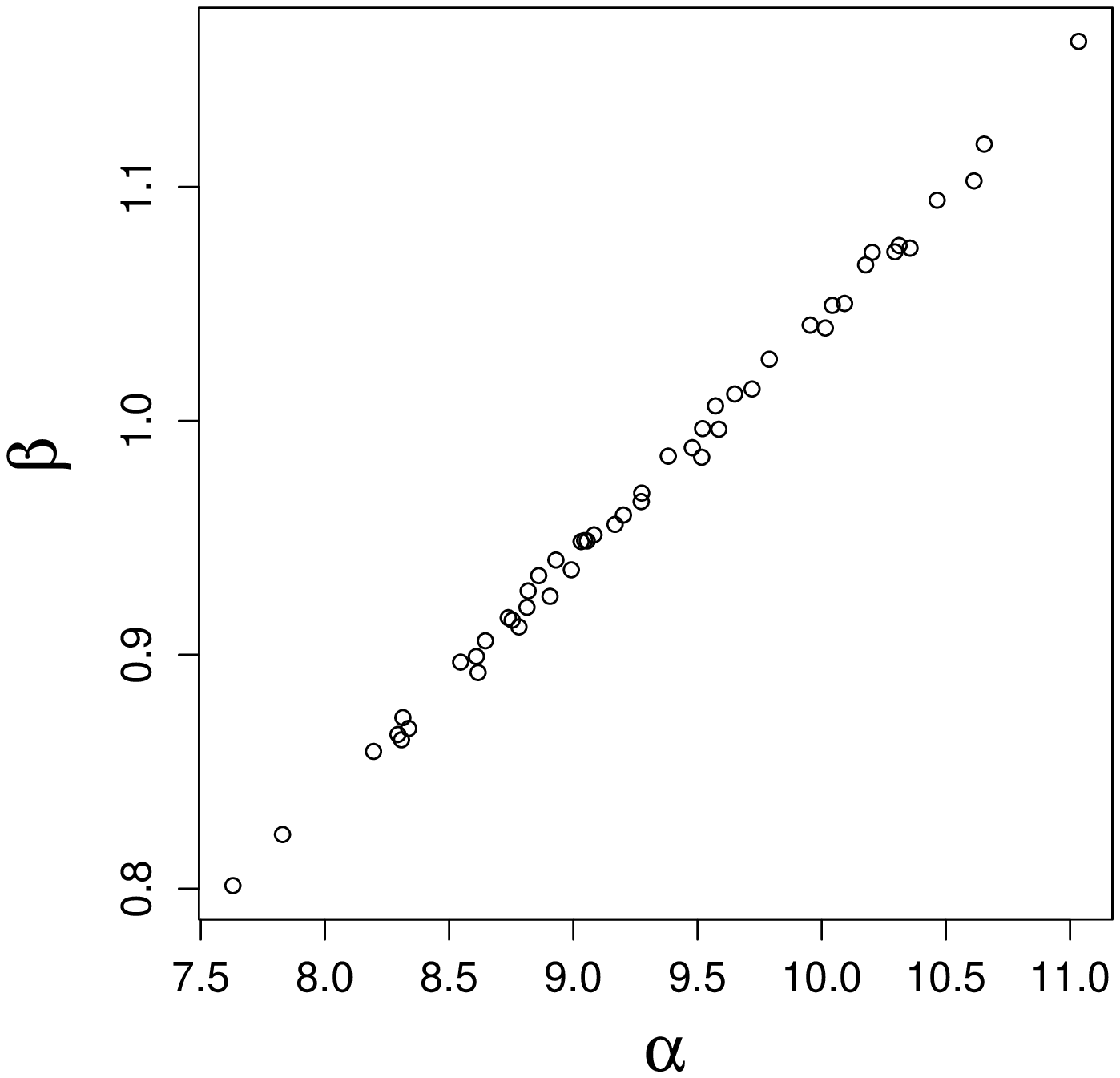, height=5cm, width=6cm} & %
\psfig{figure=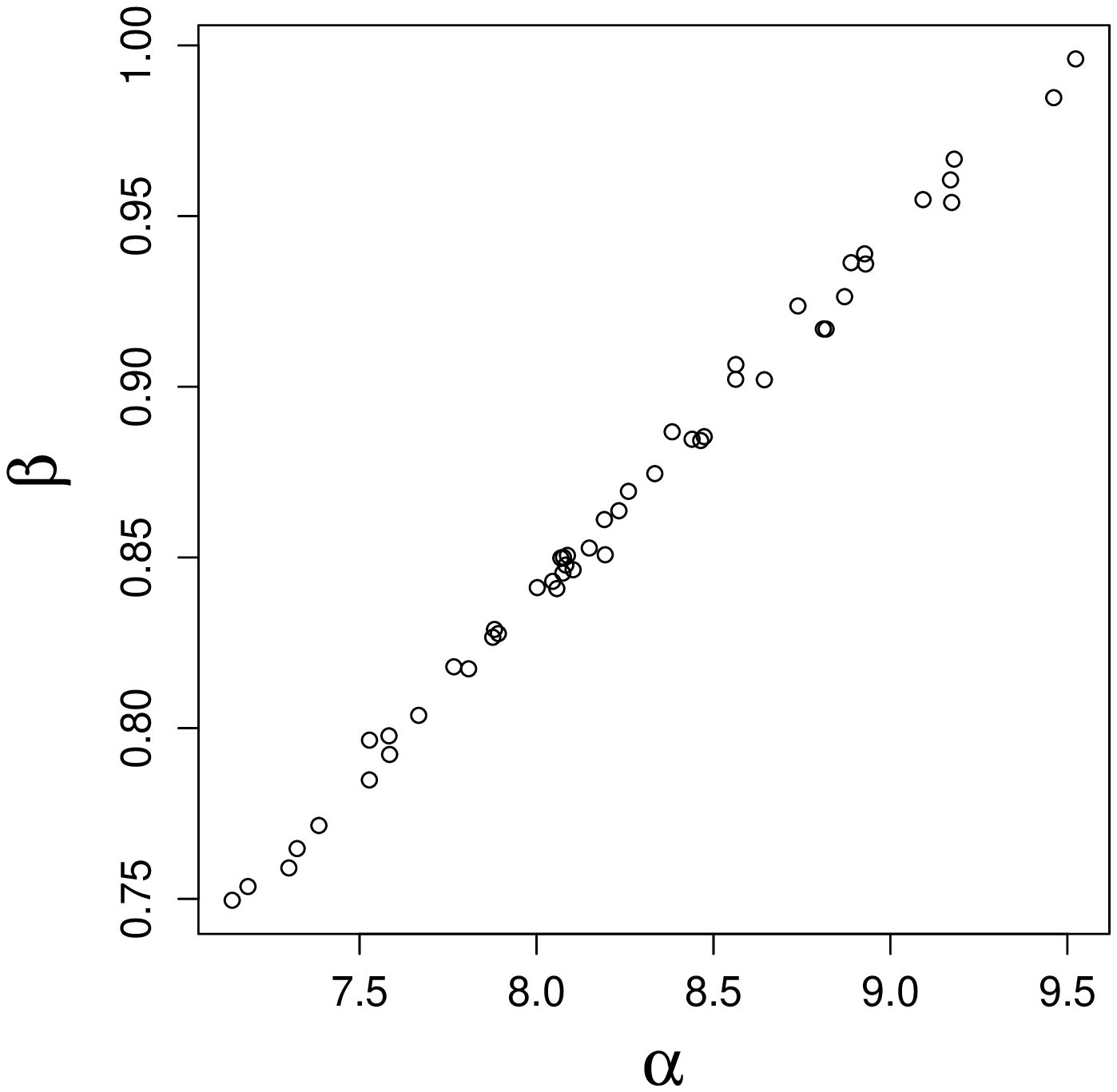, height=5cm, width=6cm} \\
(a) & (b) %
\end{tabular}
\end{center}
\caption[]{ \small AMLE estimators of $\beta$ \emph{vs.} AMLE estimators of $\alpha$: (a) 8 summary statistics; (b) 9 summary statistics.}
\label{fig:scatterab}
\end{figure}

In order to make inference on the parameters $(N,\alpha,\beta)$ using the AMLE
approach, we implement two ABC samplers using the priors
$N\sim\mbox{Unif}\{1,2,3,4,5\}$, $\alpha\sim\mbox{Unif}(5,15)$,
$\beta\sim\mbox{Unif}(0.25,1.5)$, tolerances
$\varepsilon\in\{0.5,0.4,0.3,0.2,0.15\}$ and two sets of summary
statistics. The first set of summary statistics, proposed in \cite{CK12} and
\amj{subsequently} used in \cite{M12}, consists of the mean rate of occurrence, the
coefficient of variation of the intervals between successive points, the sum
of the first five autocorrelations of the intervals, the mean of the
intervals, and the Poisson indexes of dispersion, variance divided by mean,
for intervals of length 1, 5, 10 and 20. \cite{CK12} mention that summary
statistics based on the intervals between successive points are likely to be
useful when $N$ is small, therefore we consider a second set of summary
statistics by adding a ninth quantity based on the third moment: the sample
skewness of the intervals between successive points
$\sum_{j=1}^n(x_j-\bar{x})^3/(\sum_{j=1}^n(x_j-\bar{x})^2/n)^{3/2}$. Note
that, unlike \cite{CK12} and \cite{M12}, we are taking the discrete nature of
the parameter $N$ into account. The AMLE approach is still applicable in this
context given that the maximisation of the joint posterior distribution of
$(N,\alpha,\beta)$ can be conducted by conditioning on $N$. \amj{We also
  considered a continuous prior, uniform over $[1,5]$ and obtained comparable
  results (not shown) -- although, naturally, by using a discrete prior on $N$
  instead of a continuous one, the uncertainty in the estimation of $\alpha$
  and $\beta$ is reduced. Although allowing $N$ to take a continuous range of
  values leads to an analysis which is arguably more immediately comparable to
  those presented previously in the literature, we prefer to restrict $N$ to a
  discrete set as this is consistent with the statistical interpretation of
  the parameter and the possibility of doing so is a clear advantage of the
  AMLE methodology.} Figure
\ref{fig:effectepsRP} shows the Monte Carlo variability, estimated by using
$50$ AMLE samples, for each of the two AMLE approaches based on ABC samples of
size $5000$. We can notice that the precision in the estimation of
$(\alpha,\beta)$ increases faster, as the tolerance decreases, when using 9
summary statistics. We can observe the same phenomenon from Table
\ref{table:Nest} in the estimation of $N$. \amj{(Note that the horizontal line
shows the parameters used to generate the data \emph{not} the true value of
the MLE).} {The uncertainty in the estimation of $\alpha$ and $\beta$ using the AMLE approach with the set of 9 summary statistics seems to be qualitatively comparable with that in \cite{CK12} for a small tolerance $\varepsilon$}. {Figure \ref{fig:scatterab} shows scatter plots of the AMLE estimators of $\beta$ and $\alpha$
for $\varepsilon=0.15$ and both sets of summary statistics.} \amj{This scatterplot demonstrates that the mean ($\alpha/\beta$) of the gamma distribution is much more tightly constrained by the data than the shape parameter, leading to a nearly-flat ridge in the likelihood surface. The variability in the estimated value of $\alpha/\beta$ is, in fact, rather small; while the variability in estimation of the shape parameter reflects the lack of information about this quantity in the data and the consequent flatness of the likelihood surface.}

\begin{table}[!h]
\begin{center}
\begin{tabular}[h]{|c|c|c|c|c|c|c|c|c|}
\hline
 & \multicolumn{4}{c|}{8 summary statistics} &  \multicolumn{4}{c|}{9 summary statistics}\\
\hline
$\varepsilon$ & 1 & 2 & 3 & 4 & 1 & 2 & 3 & 4 \\
\hline
$0.5$  & 29 & 0 & 1 & 20 & 33 & 0  & 15 & 2 \\
\hline
$0.4$  & 5 & 0 & 35 & 10 & 0 & 0  & 50 & 0\\
\hline
$0.3$   & 0 & 4 & 46 & 0 & 0 & 37 & 13 & 0\\
\hline
$0.2$  & 0 & 50 & 0 & 0 & 0 & 50  & 0 & 0\\
\hline
$0.15$  & 0 & 50 & 0 & 0 & 0 & 50  & 0 & 0\\
\hline
\end{tabular}
\caption{\small Replicate study with a single data realisation. Estimators of $N$ for different values of $\varepsilon$}
\label{table:Nest}
\end{center}
\end{table}

\pagebreak

\begin{figure}[h!]
\begin{center}
\begin{tabular}{cc}
\psfig{figure=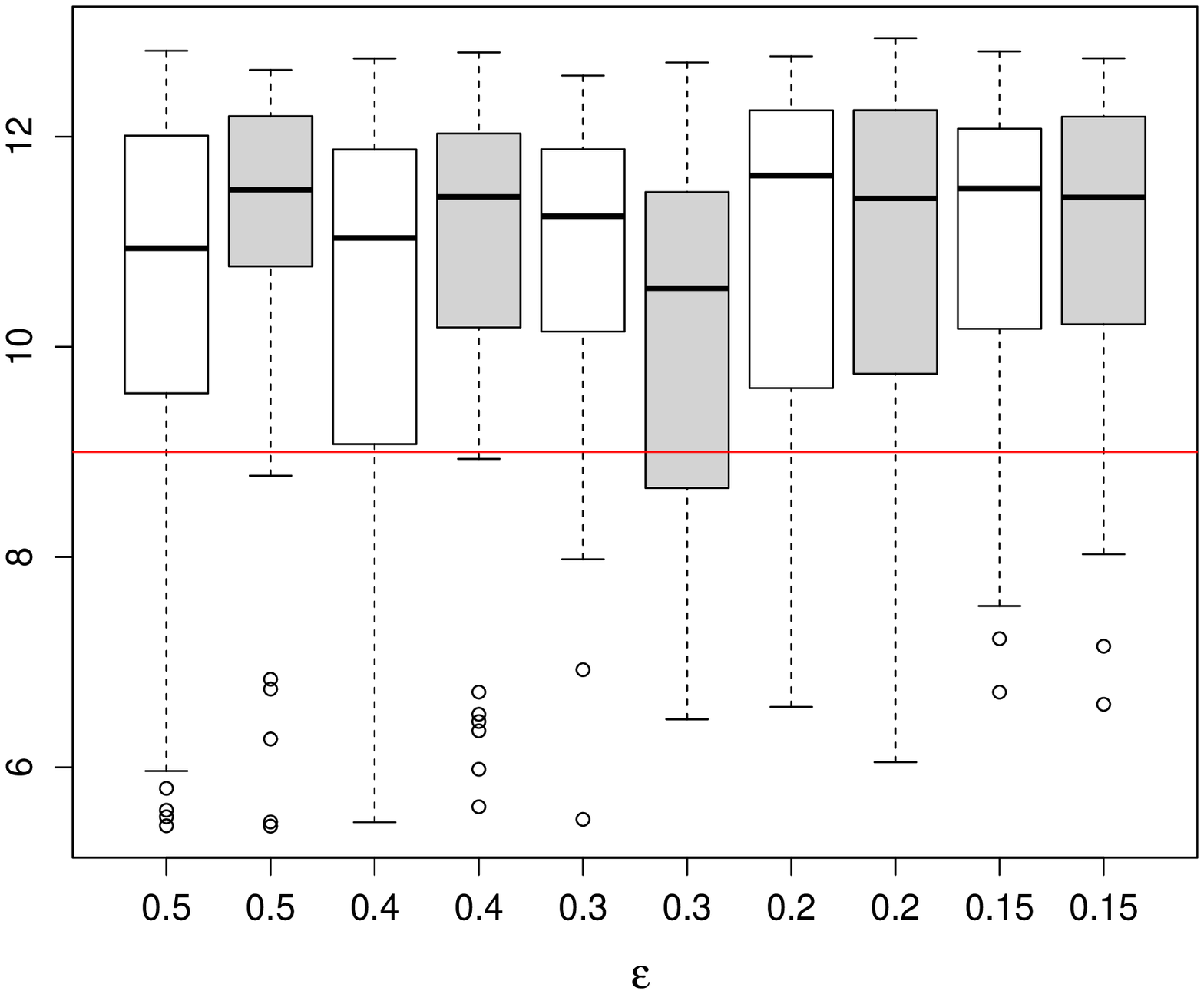, height=5cm, width=6cm} & %
\psfig{figure=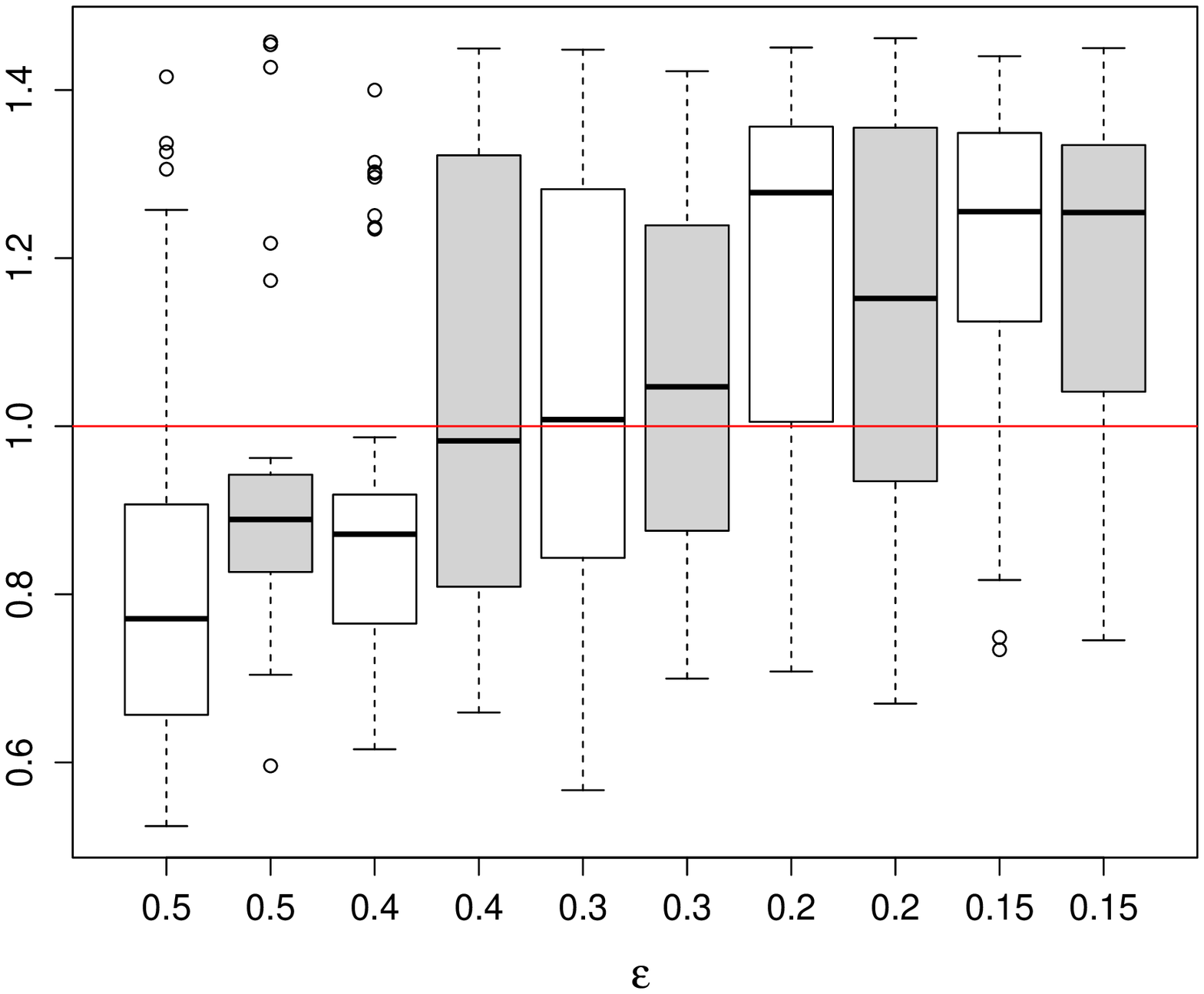, height=5cm, width=6cm} \\
(a) & (b) %
\end{tabular}
\end{center}
\caption[]{ \small Effect of $\varepsilon \in \{0.5,0.4,0.3,0.2,0.15\}$ for $n=5000$: (a) $\alpha$; (b) $\beta$. The AMLE samples with $8$ and $9$ summary statistics are presented in white and gray boxplots, respectively. The continuous red line represents the true value of the parameter.}
\label{fig:effectepsRPD}
\end{figure}

\amj{To show the variability of the estimator with different data, we
  also} compare the variability of the estimators obtained using 50 different data sets. For each data set we obtain the corresponding AMLE of $(N,\alpha,\beta)$ by using the priors $N\sim\mbox{Unif}\{1,2,3,4,5\}$, $\alpha\sim\mbox{Unif}(5,13)$ and $\beta\sim\mbox{Unif}(0.5,1.5)$, tolerances $\varepsilon\in\{0.5,0.4,0.3,0.2,0.15\}$ and the two sets of summary statistics mentioned above. Figure \ref{fig:effectepsRPD} shows the boxplots of the AMLEs for $(\alpha,\beta)$ obtained using ABC samples of size $5000$. We can observe that the behaviour of the estimators of $(\alpha,\beta)$ is fairly similar for both sets of summary statistics. Table \ref{table:Nest} also suggests an improvement in the estimation of $N$ produced by the inclusion of the sample skewness.

\begin{table}[!h]
\begin{center}
\begin{tabular}[h]{|c|c|c|c|c|c|c|c|c|}
\hline
 & \multicolumn{4}{c|}{8 summary statistics} &  \multicolumn{4}{c|}{9 summary statistics}\\
\hline
$\varepsilon$ & 1 & 2 & 3 & 4 & 1 & 2 & 3 & 4 \\
\hline
$0.5$  & 7 & 0 & 24 & 19 & 5 & 0 & 43 & 2 \\
\hline
$0.4$  & 8 & 0 & 41 & 1 & 3 & 0 & 24 & 23 \\
\hline
$0.3$   & 1 & 27 & 22 & 0 & 0 & 43 & 7 & 0 \\
\hline
$0.2$  & 0 & 46 & 4 & 0 & 0 & 44  & 6 & 0\\
\hline
$0.15$  & 0 & 47 & 3 & 0 & 0 & 46  & 4 & 0 \\
\hline
\end{tabular}
\caption{\small Replicate study with 50 data realisations. Estimators of $N$ for different values of $\varepsilon$}
\label{table:NDest}
\end{center}
\end{table}

\pagebreak

\section{Discussion}\label{sec:conclusion}
This paper presents a simple algorithm for conducting maximum likelihood
estimation via simulation in settings in which the likelihood cannot (readily)
be evaluated and provides theoretical and empirical support for that
algorithm. This adds another tool to the ``approximate computation''
toolbox. This allows the (approximate) use of the MLE in most
settings in which ABC is possible: desirable both in itself and because it is
unsatisfactory for the approach to inference to be dictated by computational
considerations. Furthermore, even in settings in which one wishes to adopt a
Bayesian approach to inference it may be interesting to obtain also a
likelihood-based estimate as agreement or disagreement between the approaches. Naturally, both ABC and AMLE being based upon the
same approximation, the difficulties and limitations of ABC are largely
inherited by AMLE. Selection of statistics in the case in which sufficient
statistics are not available remains a critical question. There
has been considerable work on this topic in recent years (see e.g. \citealp{Fear12}).

A side-effect of the AMLE algorithm is an approximate characterisation of the
likelihood surface, or in Bayesian settings of the posterior surface.
We would strongly recommend that this surface be inspected whenever
ABC or related techniques are used as even in settings in which the original
likelihood contains strong information about the parameters it is possible for
a poor choice of summary statistic to lead to the loss of this
information. Without explicit consideration of the approximation, perhaps
combined with prior sensitivity analysis, this type of issue is difficult to detect.


\section*{Acknowledgements}
AMJ gratefully acknowledges support from EPSRC grant EP/I017984/1. FJR acknowledges support from Conacyt, M{\'e}xico.

\singlespace


\begin{thebibliography}{9}
\small
\bibitem[Abraham et al.(2003)]{A03} Abraham, C., Biau, G. and Cadre, B. (2003). Simple estimation of the mode of a multivariate density. {\sl The Canadian Journal of Statistics} 31: 23--34.
\bibitem[Beaumont et al.(2002)]{B02} Beaumont, M. A., Zhang, W. and Balding, D. J. (2002). Approximate Bayesian computation in population genetics. {\sl Genetics} 162: 2025--2035.
\bibitem[Besag(1975)]{Besag75} Besag, J. (1975). Statistical Analysis of
  Non-Lattice Data. {\sl The Statistician} 24:179--195.
\bibitem[Bickel and Fr\"{u}wirth(2006)]{BF06} Bickel, D. R. and Fr\"{u}wirth, R. (2006). On a fast, robust estimator of the mode: Comparisons to other robust estimators with applications . {\sl Computational Statistics \& Data Analysis} 50: 3500--3530.
\bibitem[Cox and Kartsonaki(2012)]{CK12} Cox, D. R. and Kartsonaki, C. (2012). The fitting of complex parametric models. {\sl Biometrika} 99: 741--747.
\bibitem[Cox and Reid(2004)]{CR04} Cox, D. R. and Reid, N. (2004). A note on pseudolikelihood constructed from marginal densities. {\sl Biometrika} 91: 729--737.
\bibitem[Cox and Smith(1954)]{CS54} Cox, D. R. and Smith, W. L. (1954). On the superposition of renewal processes. {\sl Biometrika} 41: 91-–9.
\bibitem[Cule et al.(2010)]{C10} Cule, M. L., Samworth, R. J. and Stewart, M. I. (2010), Maximum likelihood estimation of a multi-dimensional log-concave density. {\sl Journal Royal Statistical Society} B 72: 545--600.
\bibitem[Dean et al.(2011)]{Dean11} Dean, T. A., Singh, S. S., Jasra A. and Peters G. W. (2011). Parameter estimation for hidden Markov models with intractable likelihoods. Arxiv preprint arXiv:1103.5399v1.
\bibitem[de Valpine(2004)]{dV04} de Valpine, P. (2004) Monte Carlo state space likelihoods by weighted posterior kernel density estimation. {\sl
Journal of the American Statistical Society} 99: 523–-536.
\bibitem[Didelot et al.(2011)]{Didelot11} Didelot, X., Everitt, R. G.,
  Johansen, A. M. and Lawson, D. J. (2011). Likelihood-free estimation of
  model evidence. {\sl Bayesian Analysis} 6: 49--76.
\bibitem[Diggle and Gratton (1984)]{Diggle84} Diggle, P. J. and Gratton,
  R. J. (1984) Monte Carlo Methods of Inference for Implicit Statistical
  Models. {\sl Journal of the Royal Statistical Society. Series B
    (Methodological)} 46:193--227.
\bibitem[Duong and Hazelton(2005)]{D05} Duong, T. and Hazleton, M. L. (2005). Cross-validation bandwidth
     matrices for multivariate kernel density estimation. {\sl Scandinavian
     Journal of Statistics} 32:485--506.
\bibitem[Duong(2011)]{D11} Duong, T. (2011). {\sl ks: Kernel smoothing}. R package version 1.8.5. http://CRAN.R-project.org/package=ks
\bibitem[Fan et al.(2012)]{F12} {Fan, Y., Nott, D. J. and Sisson, S. A. (2012). Approximate Bayesian Computation via Regression
Density Estimation. Arxiv preprint arXiv:1212.1479}
\bibitem[Fearnhead and Prangle(2012)]{Fear12} Fearnhead, P. and Prangle, D. (2012). Constructing Summary Statistics for Approximate Bayesian
  Computation: Semi-automatic ABC (with discussion). Journal of the Royal Statistical Society
  Series B (Methodology) \emph{in press}.
\bibitem[Fukunaga and Hostetler(1975)]{FH75} {Fukunaga, K. and Hostetler, L. D. (1975). The Estimation of the Gradient of a Density Function, with Applications in Pattern Recognition. {\sl IEEE Transactions on Information Theory} 21: 32--40.}
\bibitem[Gaetan and Yao(2003)]{GY03} Gaetan, C. and Yao, J. F. (2003). A
  multiple-imputation Metropolis version of the EM algorithm. {\sl Biometrika}
  90: 643--654.
\bibitem[Ionides(2005)]{Ionides05} Ionides, E. (2005). Maximum Smoothed Likelihood Estimation. {\sl Statistica Sinica} 15: 1003--1014.
\bibitem[Jaki and West(2008)]{Jaki08} Jaki, T. and West, R. W. (2008). Maximum
  Kernel Likelihood Estimation. {\sl Journal of Computational and Graphical
    Statistics} 17: 976.
\bibitem[Jing et al.(2012)]{J12} Jing, J., Koch, I. and Naito, K (2012). Polynomial
  Histograms for Multivariate Density and Mode Estimation. {\sl Scandinavian
    Journal of Statistics} 39:75--96.
\bibitem[Johansen et al.(2008)]{JDD08} Johansen, A. M., Doucet, A. and Davy, M. (2008). Particle methods for maximum likelihood parameter estimation in
  latent variable models. {\sl Statistics and Computing} 18: 47--57.
\bibitem[Konakov(1973)]{K73} Konakov, V. D. (1973). On asymptotic normality of the sample mode of multivariate distributions. {\sl Theory of Probability and its Applications} 18: 836--842.
\bibitem[Lehmann and Casella(1998)]{LC98} Lehmann, E. and Casella, G. (1998). {\sl Theory of Point Estimation} (revised edition). Springer-Verlag, New York.
\bibitem[Lele et al.(2007)]{L07} Lele, S. R., Dennis, B. and Lutscher, F. (2007). Data cloning: easy maximum likelihood estimation for complex ecological models using Bayesian Markov chain Monte Carlo methods. {\sl Ecology Letters} 10: 551--563.
\bibitem[McCulloch(1986)]{M86} McCulloch, J. H. (1986). Simple consistent estimators of stable distribution parameters. {\sl Communications in Statistics: Simulation and Computation} 15: 1109--1136.
\bibitem[Marin et al.(2011)]{Marin11} Marin, J., Pudlo, P., Robert, C. P. and Ryder, R. (2011). Approximate Bayesian Computational methods.  {\sl Statistics and Computing} \emph{in press}.
\bibitem[Marjoram et al.(2003)]{Marjoram03} Marjoram, P., Molitor, J., Plagnol, V. and Tavar{\'e}, S. (2003). Markov chain Monte Carlo without likelihoods. {\sl Proceedings of the National Academy of Sciences} USA: 15324--15328.
\bibitem[Mengersen et al.(2012)]{M12} Mengersen, K. L., Pudlo, P. and Robert, C. P. (2012). Approximate Bayesian computation via empirical likelihood. {\sl Proceedings of the National Academy of Sciences of the United States of America}, forthcoming.
\bibitem[Nolan(2001)]{N01} Nolan, J. P. (2001). Maximum likelihood estimation and diagnostics for stable distributions. In: O.E.
Barndorff-Nielsen, T. Mikosh, and S. Resnick, Eds., L{\'e}vy Processes, Birkhauser, Boston, 379--400.
\bibitem[Parzen(1962)]{P62} Parzen, E. (1962). On estimation of a probability density function and mode. {\sl Annals of Mathematical Statistics} 33: 1065--1076.
\bibitem[Peters et al.(2010)]{P10} Peters, G. W., Sisson, S. A. and Fan, Y. (2010). Likelihood-free Bayesian inference for $\alpha-$stable models. {\sl Computational Statistics \& Data Analysis} \emph{in press}. http://dx.doi.org/10.1016/j.csda.2010.10.004
\bibitem[Pritchard et al.(1999)]{Prit99} Pritchard, J. K., Seielstad, M. T., Perez-Lezaun, A., and Feldman, M. T. (1999). Population Growth of Human Y Chromosomes: A Study of Y Chromosome Microsatellites. {\sl Molecular Biology and Evolution} 16: 1791--1798.
\bibitem[Robert(2007)]{R07} Robert, C. P. (2007). The Bayesian Choice (2nd ed.). New York: Springer.
\bibitem[Robert et al.(2011)]{Robert11b} Robert, C. P., Cornuet, J., Marin, J. and Pillai, N. S. (2011). Lack of confidence in ABC model choice. {\sl Proceedings of the National Academy of Sciences of the United States of America} 108: 15112--15117.
\bibitem[Romano(1988)]{R88} Romano, J. P. (1988). On weak convergence and optimality of kernel density estimates of the mode. {\sl The Annals of Statistics} 16: 629--647.
\bibitem[Rubio and Johansen(2012)]{RJ12} Rubio, F. J. and Johansen,
  A. M. J. (March, 2012). On Maximum Intractable Likelihood. CRiSM working paper 12--04.
\bibitem[Rudin(1976)]{R76} Rudin, W. (1976). {\sl Principles of Mathematical Analysis}. New York: McGraw-Hill.
\bibitem[Sisson(2007)]{S07} Sisson, S. A., Fan, Y., and Tanaka, M. M. (2007). Sequential Monte Carlo without likelihoods. {\sl Proceedings of the National Academy of Sciences}, 104: 1760--1765.
\bibitem[Sk{\"o}ld et al.(2003)]{Skold03} Sk{\"o}ld, M. and Roberts,
  G. O. (2003). Density estimation for the Metropolis--Hastings algorithm. {\sl
    Scandinavian Journal of Statistics} 30: 699--718.
\bibitem[Tavar{\'e} et al.(1997)]{Tav97} Tavar{\'e}, S., Balding, D., Griffith, R. and Donelly, P. (1997). Inferring coalescence times from DNA sequence data. {\sl Genetics} 145: 505--518.
\bibitem[Toni et al.(2009)]{T09} Toni, T.; Welch, D.; Strelkowa, N.; Ipsen, A.; Stumpf, M.P.H. (2009). Approximate Bayesian computation scheme for parameter inference and model selection in dynamical systems. {\sl Journal of the Royal Society Interface} 6: 187--202.
\bibitem[Whitney(1991)]{W91} Whitney, K. N. (1991). Uniform Convergence in probability and stochastic equicontinuity. {\sl Econometrica} 59: 1161--1167.
\bibitem[Wilkinson(2008)]{Wilkinson08} Wilkinson, R. D. (2008). Approximate Bayesian computation (ABC) gives exact results under the assumption of error model. Arxiv preprint arXiv:0811.3355.
\bibitem[Wuertz et al.(2010)]{W10} Wuertz, D. and core team members R (2010). {\sl fBasics: Rmetrics - Markets and Basic Statistics}. R package version 2110.79.   http://CRAN.R-project.org/package=fBasics
\end{thebibliography}
\end{document}